\documentclass[acmsmall,screen]{acmart}

\usepackage[ruled,vlined,lined,commentsnumbered]{algorithm2e}
\usepackage{amsmath,amsfonts}
\usepackage{array}
\usepackage{booktabs}
\usepackage[skip=1pt,labelfont=bf]{caption}
 \usepackage{calligra}
\usepackage{color, colortbl}
\usepackage{courier}
\usepackage{csvsimple}
\usepackage{enumitem}
\usepackage{fancybox}
\usepackage{fontenc}
\usepackage{graphicx}
\usepackage{listings}
\usepackage{longtable}
\usepackage{lscape}
\usepackage{makecell}
\usepackage{marvosym}
\usepackage{moreverb}
\usepackage{multicol}
\usepackage{multirow}
\usepackage{pifont}
\usepackage{rotating}
\usepackage{setspace}
\usepackage{caption}
\usepackage{subfigure}
\usepackage[most]{tcolorbox}
\usepackage{threeparttable}
\usepackage{tikz}
\usepackage[normalem]{ulem}
\usepackage{url}
\usepackage{soul}
\usepackage{wasysym}
\usepackage{svg}
\usepackage{textcomp}
\usepackage{xcolor}
\usepackage{wrapfig}
\usepackage{pdflscape}
\usepackage{hyperref}
\usepackage{ifthen}

\usepackage{amsmath}
\usepackage{array}
\usepackage{balance}
\usepackage{microtype}
\newcolumntype{L}[1]{>{\raggedright\let\newline\\\arraybackslash\hspace{0pt}}m{#1}}
\newcolumntype{C}[1]{>{\centering\let\newline\\\arraybackslash\hspace{0pt}}m{#1}}
\newcolumntype{R}[1]{>{\raggedleft\let\newline\\\arraybackslash\hspace{0pt}}m{#1}}

\newcommand{\xy}[1]{\textcolor{black}{#1}} 
\newcommand{\toolname}{RATester\xspace}

\definecolor{darkgreen}{rgb}{114,169,119}

\newcommand{\intuition}[1]{
\begin{tcolorbox}[colback=white,boxrule=1pt,top=0pt,bottom=0pt,left=1pt,right=2pt,top=2pt,bottom=2pt]
\em #1
\end{tcolorbox}
}

\begin{document}

\title{Enhancing LLM's Ability to Generate More Repository-Aware Unit Tests Through Precise Contextual Information Injection}

\author{Xin Yin}
\affiliation{%
  \institution{The State Key Laboratory of Blockchain and Data Security, Zhejiang University}
  \city{Hangzhou}
  \country{China}
  }
\email{xyin@zju.edu.cn}

\author{Chao Ni}
\authornote{Chao Ni the corresponding author.\\
He is also with Hangzhou High-Tech Zone (Binjiang) Institute of Blockchain and Data Security.}
\affiliation{%
  \institution{The State Key Laboratory of Blockchain and Data Security, Zhejiang University}
  \city{Hangzhou}
  \country{China}
  }
\email{chaoni@zju.edu.cn}

\author{Xinrui Li}
\affiliation{%
  \institution{The State Key Laboratory of Blockchain and Data Security, Zhejiang University}
  \city{Hangzhou}
  \country{China}
  }
\email{lixinrui@zju.edu.cn}

\author{Liushan Chen}
\affiliation{%
  \institution{ByteDance Inc.}
  \city{Shenzhen}
  \country{China}
  }
\email{chenliushan@bytedance.com}

\author{Guojun Ma}
\affiliation{%
  \institution{ByteDance Inc.}
  \city{Shenzhen}
  \country{China}
  }
\email{maguojun@bytedance.com}

\author{Xiaohu Yang}
\affiliation{%
  \institution{The State Key Laboratory of Blockchain and Data Security, Zhejiang University}
  \city{Hangzhou}
  \country{China}
  }
\email{yangxh@zju.edu.cn}

\begin{abstract}

Though many learning-based approaches have been proposed for unit test generation and achieved remarkable performance, they still have limitations in relying on task-specific datasets.
Recently, Large Language Models (LLMs) guided by prompt engineering have gained attention for their ability to handle a broad range of tasks, including unit test generation.
Despite their success, LLMs may exhibit hallucinations when generating unit tests for focal methods or functions due to their lack of awareness regarding the project's global context.
These hallucinations may manifest as calls to non-existent methods, as well as incorrect parameters or return values, such as mismatched parameter types or numbers.
\xy{While many studies have explored the role of context, they often extract fixed patterns of context for different models and focal methods, which may not be suitable for all generation processes (e.g., excessive irrelevant context could lead to redundancy, preventing the model from focusing on essential information).}

To overcome this limitation, we propose \toolname, which enhances the LLM’s ability to generate more repository-aware unit tests through global contextual information injection.
To equip LLMs with global knowledge similar to that of human testers, we integrate the language server gopls, which provides essential features (e.g., definition lookup) to assist the LLM.
When \toolname encounters an unfamiliar identifier (e.g., an unfamiliar struct name), it first leverages gopls to fetch relevant definitions and documentation comments, and then uses this global knowledge to guide the LLM.
By utilizing gopls, \toolname enriches the LLM’s knowledge of the project’s global context, thereby reducing hallucinations during unit test generation.

We evaluate the effectiveness and efficiency of \toolname compared to baseline approaches by constructing a new Golang dataset from real-world projects. 
The results demonstrate the advantages of \toolname over the baselines. 
For instance, on our dataset, \toolname achieves an average line coverage of 26.25\%, representing an improvement of 16.30\% to 165.69\% over the baselines. 
Furthermore, \toolname shows superior performance in mutation testing, successfully killing 25 to 147 more mutants than the baseline approaches.
Additionally, we extend our analysis to assess the model-agnostic effectiveness of \toolname. 
These results not only confirm the effectiveness of \toolname but also underscore its universal applicability.

\end{abstract}



\maketitle

\section{Introduction}
Unit testing plays a critical role in software maintenance by enabling developers to identify defects and errors early in the development process, thereby ensuring the quality of software systems.
This not only helps lower overall product costs but also enhances developer productivity~\cite{garousi2013survey,lee2012survey,barr2014oracle}. 
Despite its significance, manually writing high-quality unit tests is both challenging and time-consuming. 
To mitigate this, a range of automated unit test generation approaches have been proposed, which can be broadly classified into three categories: traditional approaches~\cite{fraser2011evosuite,pacheco2007randoop,wang2023nxtunit}, learning-based approaches~\cite{alagarsamy2023a3test,tufano2020unit,shin2024domain,he2024unitsyn,rao2023cat}, and LLM-based approaches~\cite{ni2024casmodatest,yuan2023no,chen2024chatunitest, lemieux2023codamosa}.

Traditional approaches~\cite{fraser2011evosuite, pacheco2007randoop} primarily focus on maximizing code coverage, and research demonstrates their effectiveness in achieving high coverage~\cite{aleti2017analysing, oliveira2018mapping, panichella2015reformulating, panichella2017automated}.
Randoop~\cite{pacheco2007randoop} and EvoSuite~\cite{fraser2011evosuite} are among the most popular and widely used examples of such approaches.
Randoop, a widely recognized tool, is extensively used to generate unit tests for Java code through feedback-directed random test generation. 
EvoSuite automates the creation of test suites, aiming to maximize code coverage while minimizing test suite size and ensuring comprehensive assertions. 
Despite their success in achieving coverage goals, previous studies show that these approaches do not produce well-written, maintainable unit tests explicitly for developers to use~\cite{wang2023nxtunit, tufano2020unit}.

To overcome this limitation, learning-based approaches~\cite{alagarsamy2023a3test, tufano2020unit, shin2024domain, he2024unitsyn, rao2023cat} have gained significant attention in recent years.
AthenaTest~\cite{tufano2020unit} utilizes a Transformer-based model trained on developer-written test cases to generate accurate and readable unit tests. 
A3Test~\cite{alagarsamy2023a3test} applies domain adaptation techniques, aiming to transfer knowledge from assertion generation tasks to test case generation. 
UniTester~\cite{he2024unitsyn} leverages the UniTSyn dataset to synthesize unit tests across multiple programming languages, including Golang. 
However, these approaches treat unit test generation as a translation problem, where the goal is to translate a focal method or function into a corresponding unit test. 
They rely heavily on task-specific datasets extracted from open-source repositories.

\xy
{
To address the challenges faced by learning-based approaches, researchers are increasingly exploring pre-trained Large Language Models (LLMs) for unit test generation. 
These models generate unit tests directly from contextual information, reducing reliance on task-specific datasets by leveraging extensive pre-training on large open-source code snippets. 
Researchers~\cite{ni2024casmodatest, yuan2023no} have adopted ChatGPT to generate unit tests based on focal methods.
Despite these advancements, LLMs can still exhibit hallucinations when generating unit tests for focal methods or functions due to their lack of awareness regarding the project’s global context. 
These hallucinations can include, but are not limited to, calling non-existent methods, as well as assigning incorrect parameters and return values (e.g., mismatched parameter types or incorrect parameter numbers).
To overcome this limitation, many studies have explored the extraction of context to reduce hallucinations in the generation process of LLMs. 
ChatUniTest~\cite{chen2024chatunitest} introduces an LLM-based framework that enhances automated unit test generation with an adaptive focal context mechanism, capturing relevant context within prompts. 
It also employs a ``Generation-Validation-Repair'' process to correct errors in the generated tests.
Following that, researchers~\cite{ryan2024code, gao2025prompt,yuan2024evaluating} have explored the roles of focal context and dependency context.
These methods utilize one or more fixed patterns to extract context for the focal method: (1) focal class signature; (2) signatures of other methods and fields in the class; (3) signatures of dependent classes; and (4) signatures of dependent methods and fields in the dependent classes.
However, these fixed extraction patterns present several issues: (1) they may overlook important context; for instance, when generating a unit test for a specific focal method, the LLM might require unknown context beyond the dependencies of that focal method; (2) there is potential for redundant context, as excessive irrelevant context could lead to redundancy, preventing the model from focusing on essential information.
}

In practical development scenarios, developers are typically highly familiar with the methods, functions, and structs within the package they are working on. 
Additionally, Integrated Development Environment (IDE) tools and language servers further assist by providing information on function calls and identifier descriptions, enabling developers to produce more accurate code. 
Therefore, in this paper, we aim to provide LLMs with a project's global knowledge comparable to that of human testers by introducing \toolname, 
which enhances LLM’s ability to generate more repository-aware unit tests through global contextual information injection.
\textbf{First, we propose an LLM-based framework for unit test generation.}
LLMs are trained in an unsupervised manner using up to billions of text and code tokens. 
This extensive unsupervised learning process equips LLMs with robust reasoning capabilities, enabling them to generate unit tests without relying on task-specific training datasets. 
Therefore, we propose a novel LLM-based approach \toolname for unit test generation since the representative conversational LLM provides advanced capabilities for several tasks, including natural language processing~\cite{openai2022chatgpt}, code generation~\cite{li2023starcoder}, and unit test generation~\cite{chen2024chatunitest, ni2024casmodatest, yuan2023no}.
\textbf{Second, we develop a global-aware framework to enhance the capabilities of LLMs.}
To provide LLMs with a global knowledge similar to that of human testers, we integrate the language server gopls~\cite{gopls}, which provides features such as code completion, syntax checking, and definition lookup, significantly improving development efficiency.
When \toolname encounters unfamiliar identifiers (e.g., unfamiliar method names, unfamiliar function names, and unfamiliar struct names), it proactively invokes gopls to fetch relevant definitions and documentation comments.
By continuously leveraging the capabilities of gopls, \toolname progressively enriches the LLM's global knowledge of the project, thereby reducing hallucinations and improving the effectiveness of unit test generation.

We construct a dataset to evaluate \toolname, consisting of eight highly starred GitHub projects (with stars ranging from 29.7k to 85.5k): {beego}, {echo}, {fiber}, {frp}, {gin}, {hugo}, {nps}, and {traefik}. 
To evaluate the effectiveness and efficiency of \toolname, we compare it against five baseline approaches across three categories: one traditional approach (i.e., NxtUnit~\cite{wang2023nxtunit}), one learning-based approach (i.e., UniTester~\cite{he2024unitsyn}), and three basic LLMs (i.e., CodeLlama~\cite{roziere2023code}, DeepSeek-Coder~\cite{deepseek-coder}, and Magicoder~\cite{wei2023magicoder}). 
The results demonstrate the clear superiority of \toolname over the baselines.
For instance, on our collected dataset, \toolname achieves an average line coverage of 26.25\%, representing an improvement of 16.30\% to 165.69\% over the baselines.
Furthermore, \toolname achieves the highest performance in mutation testing, successfully killing 25 to 147 more mutants than the baseline approaches. 
We also extend our analysis to explore the model-agnostic capabilities of \toolname.
The results not only validate the effectiveness of \toolname but also emphasize its universal applicability. 
\toolname is designed to be model-agnostic, enabling it to adapt to various LLMs, further underscoring its flexibility and universality.

In summary, the key contributions of this paper include:

\textbf{A. Novel LLM-based Framework:} 
We present \toolname, an advanced LLM-based framework for unit test generation that does not rely on task-specific training datasets.
Our results demonstrate that this framework can outperform existing approaches, achieving superior performance in unit test generation.

\textbf{B. Repository-Aware Tester:}
We introduce \toolname, which utilizes the features (e.g., definition lookup) of gopls to enhance the LLM’s global knowledge of the project. 
By proactively fetching definitions and documentation comments for unfamiliar information, \toolname reduces hallucinations during unit test generation.

\textbf{C. Extensive Evaluation:} 
(1) We conduct studies on the effectiveness and efficiency of \toolname and baselines by collecting a new Golang dataset from real-world projects.
(2) We evaluate \toolname and baselines not only using compile rate and line coverage metrics but also assess their capabilities in mutation testing.
\label{sec:introduction}

\section{Motivation}
\subsection{A Motivation Example}
\begin{figure}[htbp]
    \centering
    \includegraphics[width=\linewidth]{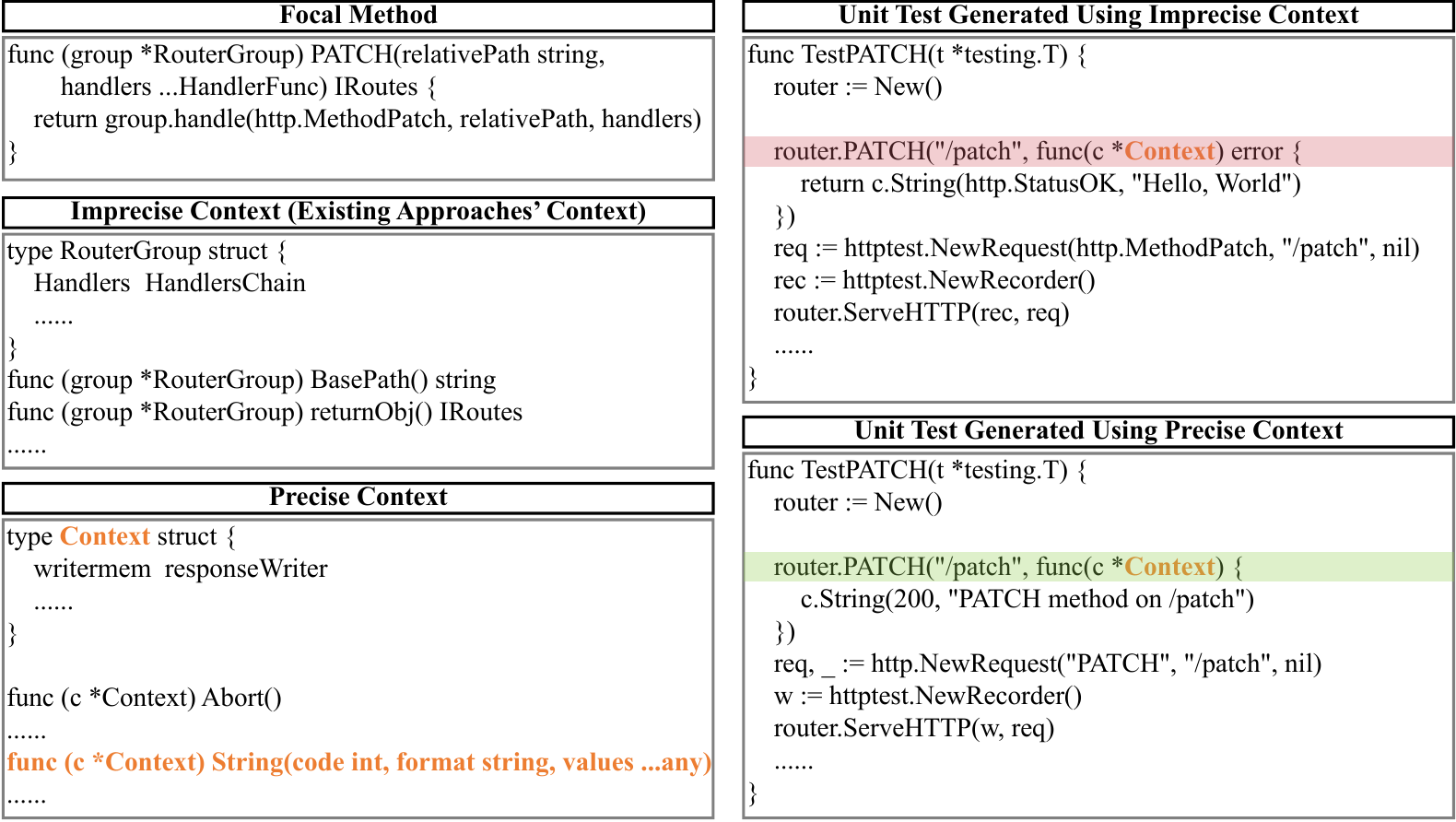}
    \caption{A focal method along with the unit tests generated by DeepSeek-Coder for a project named {gin}}
    \label{fig:motivation}
\end{figure}

Fig.~\ref{fig:motivation} shows a focal method named ``PATCH'' along with the unit tests generated by DeepSeek-Coder for a Golang project named \href{https://github.com/gin-gonic/gin}{gin}.
The upper right corner of Fig.~\ref{fig:motivation} illustrates how DeepSeek-Coder (using imprecise context) generates a unit test for the focal method without global knowledge of the project.
The unit test ``TestPATCH'' verifies whether the server can correctly handle an HTTP PATCH request sent to the ``/patch'' path and return the expected response status code of ``http.StatusOK'' and the response body ``Hello, World''.
The test creates a route instance, defines a handler for the PATCH request, and then uses the ``httptest'' package to simulate the request and capture the response, ultimately checking whether the response's status code and body meet expectations.
However, in the fourth line, this unit test encounters a compilation error: ``c.String(http.StatusOK, `Hello, World') (no value) used as value'', preventing the test from compiling. 
This issue arises because DeepSeek-Coder (using imprecise context) lacks sufficient knowledge of the project and does not know that the ``String'' method within the ``Context'' struct does not return a value, leading to hallucinations during inference.

To generate a correct unit test, one needs to understand the definition and documentation comments for the ``String'' method in the ``Context'' struct to call it correctly. 
In real-world development scenarios, human testers are typically well-acquainted with the project's global context, including the methods, functions, and structs within the package. 
IDE tools or language servers further enhance this familiarity by providing call information and identifier descriptions, aiding human testers in writing accurate code. 
As a result, human testers often refer to the tips provided by these tools to supplement their global knowledge when crafting unit tests for focal methods or functions, thereby reducing the occurrence of erroneous test cases.

\textbf{Observation.} 
\xy
{
Due to the use of fixed patterns for context extraction in existing methods, there is a certain degree of knowledge omission and redundancy in the context provided to LLMs, meaning it may not be the context that LLMs truly need during the generation process.
Given the limited input size of LLMs, it is impossible to feed all information into the model. 
Excessive irrelevant context can lead to redundancy, preventing the model from focusing on essential information, which may result in LLMs exhibiting hallucinations when generating unit tests for focal methods or functions.
These hallucinations may include invoking non-existent methods, setting incorrect parameters and return values (e.g., parameter type mismatches or incorrect number of parameters).
We should provide the model with efficient information necessary for generation, rather than fixed selections, to reduce the interference of redundant information.
}

In real-world development scenarios, human testers are typically very familiar with the package, which helps reduce the occurrence of such hallucinations.
Therefore, we also conduct a foundational experiment to simulate human testers' global knowledge of the package by providing the definition and documentation comments for the ``Context'' struct as input above the focal method.
In this case, we find that DeepSeek-Coder (using precise context) successfully generates a compilable unit test, as shown on the lower right corner of Fig.~\ref{fig:motivation}.
In the fifth line, DeepSeek-Coder (using precise context) correctly uses the ``String'' method within the ``Context'' struct without attempting to set a return value. 
We believe this results from DeepSeek-Coder's foundational understanding of the ``Context'' struct, which prevents it from generating erroneous usages.
This foundational knowledge reflects DeepSeek-Coder's understanding of the project, similar to a human tester's familiarity with it.
High-quality foundational knowledge enables the model to generate higher-quality unit tests.

\subsection{Key Ideas}
Based on the above observation, we propose \toolname for unit test generation that utilizes gopls to fetch definitions and documentation comments for unfamiliar structs, methods, functions, and more. 
\xy{This approach provides LLMs with the precise context needed for generation, effectively reducing hallucinations during unit test generation.}

\textbf{(1) LLM-based generation approach.} 
Unlike learning-based approaches (e.g., TOGA~\cite{dinella2022toga}, A3Test~\cite{alagarsamy2023a3test}, and UniTester~\cite{he2024unitsyn}), LLMs are trained in an unsupervised manner using up to billions of text and code tokens.
This large-scale unsupervised learning process allows LLMs to have strong reasoning capabilities and be applied for unit test generation without relying on training with a large amount of task-specific data.
Therefore, we propose an LLM-based generation approach, namely \toolname, since the representative conversational LLM provides advanced capabilities for several tasks~\cite{yin2024multitask,yin2024rectifier,yin2024thinkrepair,xia2023keep,xia2023automated}, including unit test generation~\cite{schafer2023empirical,ni2024casmodatest}.

\textbf{(2) Proactively fetch global contextual information.}
\xy
{
Due to the limited input size of LLMs, it is impossible to feed all information into them. 
Consequently, existing approaches use fixed patterns to select context information to include in the prompts.
However, these fixed selection approaches suffer from issues of knowledge omission and redundancy.
In this paper, to endow LLMs with a global knowledge similar to that of human testers, we introduce the Golang language server gopls, which can provide features (e.g., definition lookup) for LLMs. 
For example, as illustrated on the right side of Fig.~\ref{fig:motivation}, when \toolname encounters the unfamiliar method ``Context'', it proactively uses gopls to fetch specific definitions and documentation comments to avoid erroneous usage.
Therefore, we propose \toolname, an global-aware tester for unit test generation that utilizes gopls to fetch definitions and documentation comments for unfamiliar structs, methods, functions, and more. 
By leveraging the capabilities of gopls, we aim to enhance the LLM’s global understanding of the project and reduce hallucinations during unit test generation.
}
\label{sec:motivate}

\section{Our Approach: \toolname}
In this section, we present the methodology behind our \toolname approach. 
We begin with an overview of the approach, followed by a detailed discussion of each component.

\subsection{Overview}
\begin{figure}[htbp]
    \centering    
    \includegraphics[width=\linewidth]{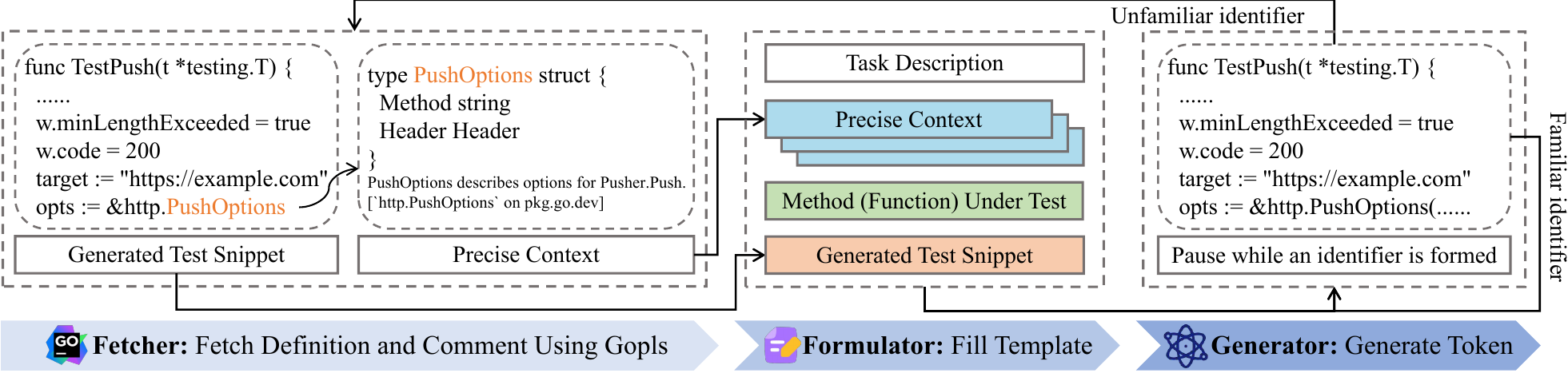}
    \caption{Overview of \toolname}
   \label{fig:overview}
\end{figure}

As shown in Fig.~\ref{fig:overview}, our approach consists of three components: Fetcher, Formulator, and Generator. 
Given the focal method or function that needs to be tested (referred to as the method or function under test), each component plays a distinct role in the unit test generation task:

\begin{itemize}[leftmargin=*]
\item \textbf{Fetcher} 
fetches the definition and documentation comment of unfamiliar identifier (e.g., struct name) using language server (e.g., gopls) according to the given information.

\item \textbf{Formulator} 
fills the fetched code context and the test snippet with newly generated identifiers into the prompt template, and then formulates them as input for the generator.

\item \textbf{Generator} 
leverages the formulated input to perform the unit test generation. 
We use DeepSeek-Coder~\cite{deepseek-coder} as the generator, which can be replaced with various LLMs (e.g., CodeLlama~\cite{roziere2023code}). 

\end{itemize}

\subsection{Fetcher}
\label{sec:fetcher}
When generating unit tests for focal methods or functions, LLMs often produce hallucinations, which can manifest as calls to non-existent methods, as well as incorrect parameter assignments and return values, such as mismatched parameter types or an improper number of parameters. 
In contrast, human testers typically possess a strong understanding of the various methods, functions, and structs within the package during the test development process. 
Additionally, IDE tools and language servers provide essential support by offering information on function calls and identifier descriptions, facilitating the creation of accurate code. 
Consequently, human testers frequently leverage insights from these tools to enhance their global knowledge while crafting unit tests, ultimately reducing the likelihood of erroneous test cases.

In this paper, \toolname serves as a Fetcher by utilizing gopls to provide LLMs with a global knowledge comparable to that of human testers. 
Gopls~\cite{gopls}, the Go language server, facilitates interactions with editors such as Visual Studio Code. 
By leveraging the capabilities of gopls, we can improve the LLM’s global understanding of the package by providing precise context, thereby mitigating hallucinations during the generation of unit tests.

In the initial stage of unit test generation, \toolname actively queries gopls for the definitions and documentation comments of the receiver type (which does not exist if a function is being tested), the parameter types, and the return type of the focal method or function.
All fetched information is then filled into the prompt template.
During the continuous phase of unit test generation, whenever the LLM generates an unfamiliar identifier (e.g., new function name and new struct name), \toolname proactively utilizes gopls to check whether the identifier exists in the current package and fetches its definition and documentation comments. 
After this, the fetched information is also filled into the prompt template by the Formulator (refer to Section~\ref{sec:formulator} for more details).
By leveraging gopls to proactively fetch definitions and documentation comments at both stages and enriching the prompt with this knowledge, the LLM gains a comprehensive understanding of unfamiliar information. 
This process closely resembles how human testers utilize IDE tools to look up definitions and documentation for unfamiliar methods, functions, and structs. 
Our \toolname approach enables the LLM to act as a tester with extensive project knowledge, thereby enhancing its testing capabilities.

\begin{figure}[htbp]
    \centering    
    \includegraphics[width=\linewidth]{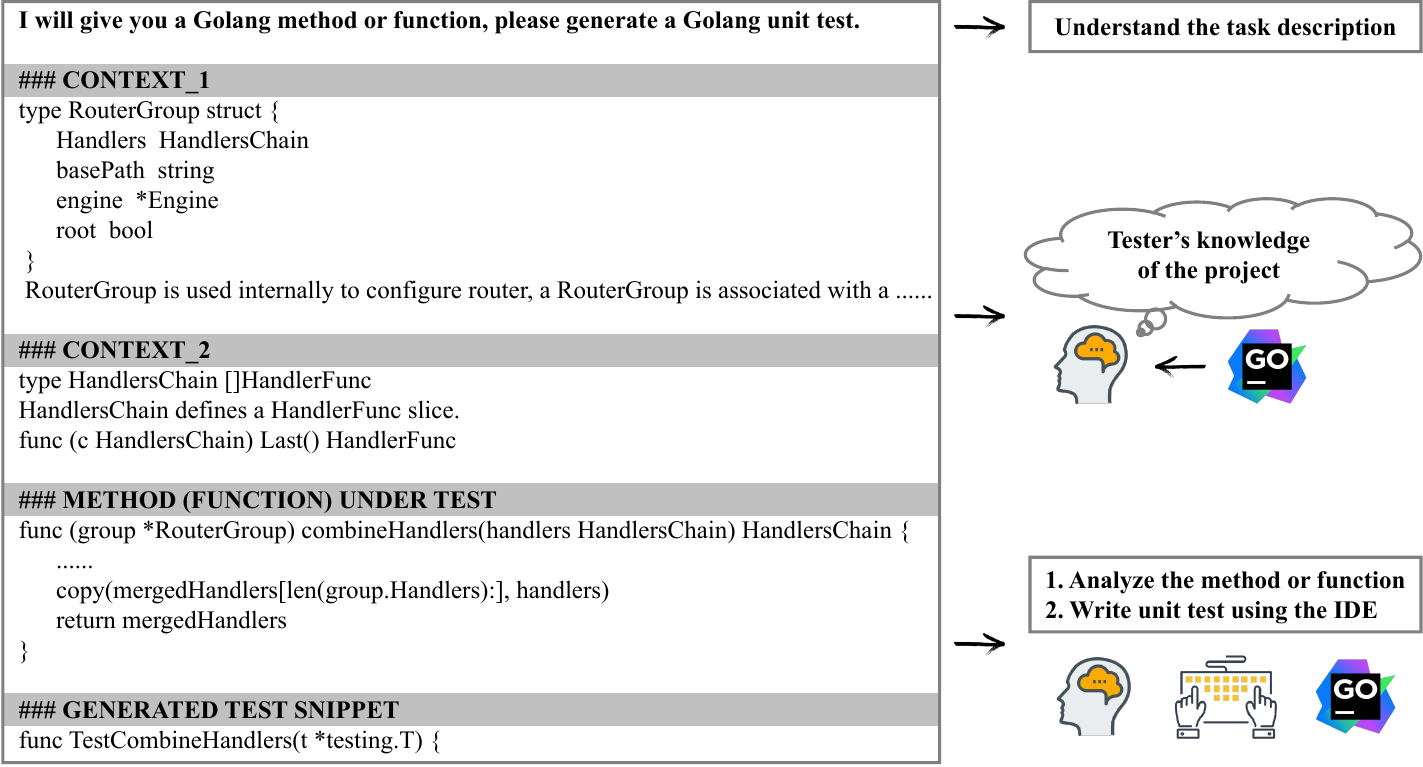}
    \caption{An example of prompt for the unit test generation}
   \label{fig:prompt}
\end{figure}

\subsection{Formulator}
\label{sec:formulator}
In both the initial stage and the continuous phase of unit test generation, the formulator fills the fetched code context and the test snippet into the prompt template, where fill means the procedure of inserting content into corresponding positions of the template.
As shown in Fig.~\ref{fig:prompt}, this prompt template consists of four main parts:

\begin{itemize}[leftmargin=*]

\item \textbf{Task Description.}
\toolname provides the LLM with the description constructed as \textit{``I will give you a Golang method or function, please generate a Golang unit test''}. 
This part aids the LLM in understanding the task description, simulating the process by which a human tester comprehends the objectives of the task.

\item \textbf{Precise Context.} 
\toolname provides the fetched definitions and documentation comments to LLM.
The code context continuously expands as the generation process progresses, enhancing the LLM's global knowledge of the project.
This part simulates the global project knowledge that human testers possess with the assistance of IDE tools.

\item \textbf{Method (Function) Under Test.}
\toolname provides the focal method or function to LLM.
We also prefix the focal method or function with \textit{``\#\#\# METHOD (FUNCTION) UNDER TEST''} to directly indicate LLM about the context of the method or function.
This part simulates the scenario in which human testers review the method or function being tested.

\item \textbf{Generated Test Snippet}.
\toolname provides the LLM with the unit test generated in the previous round, along with a new identifier.
In the initial stage, the generated test snippet is explicitly set as \textit{``func Test\{name\}(t *testing.T)''}.
As this part continues to expand, it simulates the iterative process of human testers writing unit tests.

\end{itemize}

\IncMargin{-0.5em}
\begin{algorithm} 

    \textbf{Input: }Focal method (function) ${M}$, Focal method (function) name ${N}$, Prompt template ${P}$;
    
    \textbf{Initialize: }list\_i = [], token\_length = 0, test\_snippet = ``func Test\{${N}$\}(t *testing.T) \{'';
    
    \emph{1. Fetch definitions and documentation comments of ${M}$ (Section~\ref{sec:fetcher});}

    \emph{2. Append definitions and documentation comments to code context;}

    \emph{3. Fill code context and test snippet into ${P}$ (Section~\ref{sec:formulator});}

    \While{token\_length++ < 512} {
        \emph{Generate token t using ${P}$;}

        \emph{test\_snippet += t;}
        
        \If{is\_golang\_identifier\_part(t)} {           
            \emph{Append t to list\_i;}
        }
        \Else{
        
            \If{list\_i is not empty} {          
            
                \emph{identifier = concatenate(list\_i);}
                
                \If{identifier not in code context} {
                    \emph{1. Fetch definitions and documentation comments of identifier;}

                    \emph{2. Append definitions and documentation comments to code context;}
                }
                \emph{Set list\_i to empty;}
            }
            
        }

        \emph{Fill code context and test snippet into ${P}$;}
    }

    \textbf{Output: }All generated tokens;
    \caption{Unit Test Generation Process}
    \label{alg}
\end{algorithm}

\subsection{Generator}
The generator leverages results returned from the formulator and performs the tasks of unit test generation accordingly. 
It continually generates tokens until a complete identifier is formed.
For unfamiliar identifiers, \toolname actively invokes the Fetcher to supplement the code context, while for familiar identifiers, \toolname fills the new identifier into the prompt to generate next token.
As shown in Algorithm~\ref{alg}, \toolname initializes a List: list\_i to store the currently generated identifiers.
If the token being generated can be part of an identifier (i.e., it follows Golang's identifier naming rules), it is added to list\_i. 
If the token cannot be part of an identifier (e.g., the LLM generates a character like `.'), and list\_i is not empty, the process considers that a complete identifier has been generated.
If the generated identifier is not found in the code context, \toolname uses gopls to fetch the identifier's definition and documentation comments, which are then filled into the prompt template for the next generation step (refer to Section~\ref{sec:fetcher} and Section~\ref{sec:formulator} for more details).
By continuously leveraging gopls to fetch the code context, the LLM acquires sufficient global knowledge, thereby reducing the likelihood of hallucinations during the generation process.
In this paper, we adopt DeepSeek-Coder~\cite{deepseek-coder} as the backend LLM.
\toolname is flexible to include other LLMs as the backend model (e.g., CodeLlama~\cite{roziere2023code} and Magicoder~\cite{wei2023magicoder}).
\label{sec:approach}

\section{Experimental Design}
\label{sec:setup}
In this section, we first present our collected dataset and then introduce the baseline approaches.
Following that, we describe the performance metrics as well as the experimental setting.

\begin{table}[htbp]
  \centering
  \caption{The statistic of constructed dataset}
  \resizebox{.55\linewidth}{!}
  {
    \begin{tabular}{l|crr}
    \toprule
    \textbf{Project} & \textbf{Star} & 
    \makecell{\textbf{Focal Method}\\\textbf{and Function (\#)}} & 
    \makecell{\textbf{Line Coverage}\\\textbf{of Unit Tests (\%)}} \\
    \midrule
    \textbf{beego} & 31.5k & 2,688 & 38.78\% \\
    \textbf{echo} & 29.7k & 419 & 93.58\% \\
    \textbf{fiber} & 33.6k & 765 & 85.46\% \\
    \textbf{frp} & 85.5k & 864 & 2.59\% \\
    \textbf{gin} & 78.5k & 449 & 95.53\% \\
    \textbf{hugo} & 75.4k & 3,829 & 76.54\% \\
    \textbf{nps} & 30.6k & 455 & 0.51\% \\
    \textbf{traefik} & 50.9k & 1,726 & 58.95\% \\
    \bottomrule
    \end{tabular}%
  }
  \label{tab:dataset}%
\end{table}%

\subsection{Dataset Construction}
We construct a dataset to evaluate \toolname, consisting of eight highly-starred GitHub projects (with stars ranging from 29.7k to 85.5k): \href{https://github.com/beego/beego}{beego}, \href{https://github.com/labstack/echo}{echo}, \href{https://github.com/gofiber/fiber}{fiber}, \href{https://github.com/fatedier/frp}{frp}, \href{https://github.com/gin-gonic/gin}{gin}, \href{https://github.com/gohugoio/hugo}{hugo}, \href{https://github.com/ehang-io/nps}{nps}, and \href{https://github.com/traefik/traefik}{traefik}. 
Since \toolname focuses on generating unit tests for focal methods and functions, we extract all methods and functions from each project. 
The detailed information for each project is displayed in Table~\ref{tab:dataset}.
In addition to the star count and the number of successfully extracted focal methods and functions, we also run all unit tests within the projects and show the line coverage.

\subsection{Baselines}

To investigate the effectiveness of \toolname, we consider six baselines for a comprehensive comparison, including one traditional approach, one learning-based approach, and four LLM-based approaches:

\textbf{Traditional Approach.}
To present the traditional approach, we employ NxtUnit~\cite{wang2023nxtunit}, an automatic unit test generation tool for Go that leverages random testing and is particularly suited for microservice architectures.
It offers three types of interfaces: an integrated development environment (IDE) plugin, a command-line interface (CLI), and a web-based platform.
NxtUnit's random-based strategy allows it to quickly generate unit tests, making it ideal for smoke testing and rapid quality feedback.
However, NxtUnit may sometimes fail to generate test cases due to issues like compilation errors or test crashes.
As a result, NxtUnit only provides test cases that can be executed successfully.

\textbf{Learning-based Approach.} 
To present the learning-based approach, we utilize the transformer-based generation model, UniTester~\cite{he2024unitsyn}. 
This model is trained on the UniTSyn dataset and is capable of synthesizing unit tests for programs in multiple languages, including Golang.
As the published code for UniTester lacks the model checkpoint, we retrain the UniTester model following the settings described in the paper and using the UniTSyn dataset. 
To prevent data leakage, we exclude projects from the training set that overlap with those in our collected dataset.

\textbf{LLM-based Approach.}
To represent the LLM-based approach, we utilize ChatUniTest~\cite{chen2024chatunitest} and three LLMs to generate unit tests for each focal method and function without fine-tuning.
The models we select are recently released: CodeLlama~\cite{roziere2023code}, DeepSeek-Coder~\cite{deepseek-coder}, and Magicoder~\cite{wei2023magicoder}.

\begin{itemize}[leftmargin=*]
    \item
    \textbf{CodeLlama} proposed by Rozière et al.~\cite{roziere2023code} is a collection of large pre-trained language models for code, built on Llama 2 architecture.
    These models achieve state-of-the-art among open-source models for code-related tasks, offer infilling capabilities, support for large input contexts, and robust zero-shot instruction-following abilities for programming problems.

    \item 
    \textbf{DeepSeek-Coder} developed by DeepSeek AI~\cite{deepseek-coder}  
    comprises a series of code language models, each trained from scratch on 2 trillion tokens.
    The training data includes 87\% code and 13\% natural language, covering both English and Chinese.
    DeepSeek-Coder demonstrates state-of-the-art performance among open-source code models, excelling across multiple programming languages and various benchmarks.

    \item 
    \textbf{Magicoder} proposed by Wei et al.~\cite{wei2023magicoder} is trained on 75K synthetic instruction data using OSS-Instruct, a novel approach that leverages open-source code snippets to generate diverse instruction data for code-related tasks. 
    The approach aims to mitigate the inherent bias in LLM-generated synthetic data by harnessing the vast resources of open-source code, leading to more realistic and controllable data generation.

\end{itemize}

\subsection{Evaluation Metrics} 
To evaluate the performance of \toolname and baseline approaches, we use Compile Rate and Line Coverage as primary metrics:

\textbf{Compile Rate} represents the proportion of test cases that can be successfully compiled and executed out of the total number generated. 
A higher compile rate reflects better quality and reliability in the generated test cases.

\textbf{Line Coverage} quantifies the percentage of code lines executed by the test cases, offering insights into the effectiveness of the tests in covering different parts of the code. 
A higher line coverage indicates that a larger portion of the code is being tested.

While Compile Rate and Line Coverage are valuable, they do not fully assess test quality.
To provide a more comprehensive evaluation of the unit tests generated by \toolname, we also employ mutation testing. 
We use Gremlins~\cite{gremlins} to introduce mutations into the projects and evaluate the number of mutants killed by the unit tests, along with mutator coverage.

\subsection{Implementation Details}
We develop the unit test generation in Python, utilizing PyTorch~\cite{pytorch} implementations of LLMs (i.e., CodeLlama 7B, DeepSeek-Coder 6.7B, and Magicoder 6.7B).
We use the Hugging Face API~\cite{huggingface} to load the model weights and generate outputs.
We also adhere to the best-practice guide~\cite{shieh2023best} for our prompt.
For the baseline comparisons, we directly use the settings provided in the NxtUnit's~\cite{wang2023nxtunit} original paper to generate unit tests.
Since the published code for UniTester~\cite{he2024unitsyn} does not include the model checkpoint, we retrain the UniTester model using the settings and UniTSyn dataset provided in the original paper. 
To avoid data leakage, we exclude any data from the UniTSyn dataset that overlaps with those in our collected dataset during training.
Considering both the performance improvements and the associated generation costs, we generate one unit test for each focal method and function (refer to Section~\ref{sec:rq3} for more details) and test them using the go test command.
Our evaluation is conducted on a 32-core workstation equipped with an Intel(R) Xeon(R) Platinum 8358P CPU @ 2.60GHz, 2TB RAM, and 8×NVIDIA A800 80GB GPU, running Ubuntu 20.04.6 LTS.
\label{sec:experiment}

\section{Experimental Results}
To investigate the effectiveness of \toolname on unit test generation, our experiments focus on the following three research questions:

\begin{itemize}[leftmargin=*]
\item \textbf{RQ-1 Effectiveness Comparison.} {\em How does the performance of \toolname compare with the baselines in unit test generation?}

\item \textbf{RQ-2 Model-Agnostic Analysis.} {\em 
What are the model-agnostic capabilities of \toolname in unit test generation?}

\item \textbf{RQ-3 Efficiency Comparison.} {\em How does the efficiency of \toolname compare with the baselines in unit test generation?}
\end{itemize}

\subsection{RQ-1: Effectiveness of \toolname}
\label{sec:rq1}
\noindent
\textbf{Objective.}
To reduce the hallucination issues that LLMs experience during unit test generation (e.g., invoking non-existent methods and setting incorrect parameters and return values), we propose the \toolname approach.
This approach utilizes the definition lookup feature provided by gopls to dynamically fetch relevant code context during the generation process. 
By supplying LLMs with more project-specific knowledge, we aim to minimize hallucinations.
In this section, our objective is to investigate whether \toolname outperforms previous unit test generation approaches in terms of effectiveness.

\noindent
\textbf{Experimental Design.}
In this RQ, we employ DeepSeek-Coder as the backend model for \toolname. 
To facilitate a fair comparison, we consider three baselines: NxtUnit~\cite{wang2023nxtunit}, UniTester~\cite{he2024unitsyn}, and ChatUniTest~\cite{chen2024chatunitest}. 
For NxtUnit, we use the default settings from the original paper. For ChatUniTest, we utilize DeepSeek-Coder as the backend model.
As the published code for UniTester lacks the model checkpoint, we retrain the UniTester model using the settings and UniTSyn dataset from the original paper.
To prevent data leakage, we exclude any overlapping data between the UniTSyn and our collected datasets during training.

For a comprehensive performance comparison between the baselines and \toolname, we conduct two distinct experiments across eight real-world projects. 
The first experiment involves executing all generated unit tests within each project, recording the compile rate and line coverage.
In the second experiment, we extend our evaluation with mutation testing, using Gremlins~\cite{gremlins} to mutate the projects and measure the number of mutants killed by the generated unit tests, along with the mutator coverage. 
This experiment demonstrates the effectiveness of unit tests generated by \toolname in detecting unknown defects.

\noindent 
\textbf{Results.} We discuss the results from the aspects of compile rate, line coverage, and mutation testing, respectively.

\textbf{\underline{Effectiveness of \toolname in Compile Rate and Line Coverage.}}
Table~\ref{tab:rq1_compile_rate_and_line_coverage} shows the compile rate and line coverage of unit tests generated by \toolname and the baselines.
We observe that \toolname consistently outperforms the baselines across all projects. 
Specifically, the compile rate of unit tests generated by \toolname significantly improves from 16.67\%–63.56\% to 45.58\%–69.49\%, with the average compile rate increasing from 24.61\% to 61.84\%.
Note that NxtUnit only provides test cases that can be executed successfully. 
Therefore, we do not record the compile rate for the test cases generated by NxtUnit.

In addition to the compile rate, \toolname demonstrates a significant enhancement in line coverage. 
Across all evaluated projects, \toolname increases the line coverage from a range of 7.49\%–53.92\% to 12.92\%–58.09\%.
This results in an average improvement of 16.30\%–165.69\% when compared to the baseline approaches. 
Such a substantial increase in line coverage indicates that \toolname is more effective in generating comprehensive unit tests, thereby enhancing the overall robustness of the tested software.

\begin{table}[htbp]
    \centering
    \caption{RQ-1: \toolname vs. Baselines across different projects in compile rate and line coverage}
    \resizebox{\linewidth}{!}{
        \begin{tabular}{l|rrrr|rrrr}
        \toprule
        \multirow{1.5}[4]{*}{\textbf{Projects}} & \multicolumn{4}{c|}{\textbf{Compile Rate}} & \multicolumn{4}{c}{\textbf{Line Coverage}} \\
        \cmidrule{2-9}      & \textbf{NxtUnit} & \textbf{UniTester} & \textbf{ChatUniTest} & \textbf{\toolname} & \textbf{NxtUnit} & \textbf{UniTester} & \textbf{ChatUniTest} & \textbf{\toolname} \\
        \midrule
        \textbf{beego} & - & 31.87\% & 56.38\% & 68.75\% & 20.51\% & 12.71\% & 27.85\% & 31.91\% \\
        \textbf{echo} & - & 22.36\% & 39.65\% & 45.58\% & 10.77\% & 9.89\% & 23.09\% & 24.50\% \\
        \textbf{fiber} & - & 21.22\% & 42.44\% & 59.61\% & 20.24\% & 9.33\% & 19.33\% & 26.31\% \\
        \textbf{frp} & - & 28.57\% & 52.35\% & 66.90\% & 12.84\% & 8.62\% & 13.20\% & 14.43\% \\
        \textbf{gin} & - & 32.91\% & 63.56\% & 69.49\% & 21.92\% & 11.38\% & 53.92\% & 58.09\% \\
        \textbf{hugo} & - & 24.87\% & 48.76\% & 61.51\% & 12.34\% & 9.57\% & 18.78\% & 25.02\% \\
        \textbf{nps} & - & 16.67\% & 53.11\% & 64.84\% & 16.48\% & 7.49\% & 12.66\% & 16.82\% \\
        \textbf{traefik} & - & 18.44\% & 46.25\% & 58.05\% & 10.45\% & 10.01\% & 11.69\% & 12.92\% \\
        \midrule
        \textbf{Average} & - & 24.61\% & 50.31\% & 61.84\% & 15.69\% & 9.88\% & 22.57\% & 26.25\% \\
        \bottomrule
        \end{tabular}%
    }
    \label{tab:rq1_compile_rate_and_line_coverage}%
\end{table}%

In Table~\ref{tab:rq1_enhance_the_original_unit_test}, we evaluate the ability of unit tests generated by \toolname and the baselines to enhance the line coverage of the original unit tests across the eight projects. 
The ``Original'' column represents the line coverage of the original unit tests. 
In contrast, the ``Original+NxtUnit'' column displays the total coverage achieved by combining the original tests with those generated by NxtUnit.
Similarly, the "Original+UniTester" column illustrates the total coverage obtained by integrating the original tests with those produced by UniTester.
Furthermore, the ``Original+ChatUniTest'' column reflects the total coverage resulting from the integration of the original tests with those generated by ChatUniTest. 
Finally, the ``Original+\toolname'' column indicates the total coverage obtained by combining the original tests with those generated by \toolname. 
Overall, both \toolname and the baselines successfully enhance the line coverage of the original unit tests; however, \toolname demonstrates a more significant improvement.
Specifically, the line coverage increases substantially from a range of 0.51\%–95.53\% to 14.46\%–95.57\%. 
In addition, the average compile rate also shows a notable improvement, rising from 56.49\% to 60.98\%. 
This indicates that \toolname not only enhances line coverage but also serves to complement human-written unit tests, thereby contributing to better overall software quality.

\begin{table}[htbp]
    \centering
    \caption{RQ-1: The line coverage achieved by combining the original unit tests with those generated by \toolname and the baselines}
    \resizebox{.7\linewidth}{!}{
    \begin{tabular}{l|rrrrr}
    \toprule
        \textbf{Projects} & \textbf{Original} & \makecell{\textbf{Original}+\\\textbf{NxtUnit}} &
        \makecell{\textbf{Original}+\\\textbf{UniTester}} &
        \makecell{\textbf{Original}+\\\textbf{ChatUniTest}} &
        \makecell{\textbf{Original}+\\\textbf{\toolname}} \\
        \midrule
        \textbf{beego} & 38.78\% & 38.79\% & 38.78\% & 40.93\% & 41.82\% \\
        \textbf{echo} & 93.58\% & 93.58\% & 93.59\% & 93.68\% & 93.96\% \\
        \textbf{fiber} & 85.46\% & 86.15\% & 85.56\% & 86.35\% & 86.59\% \\
        \textbf{frp} & 2.59\% & 13.51\% & 9.31\% & 14.01\% & 14.46\% \\
        \textbf{gin} & 95.53\% & 95.54\% & 95.54\% & 95.56\% & 95.57\% \\
        \textbf{hugo} & 76.54\% & 76.58\% & 76.60\% & 76.77\% & 77.20\% \\
        \textbf{nps} & 0.51\% & 16.50\% & 7.56\% & 12.29\% & 16.83\% \\
        \textbf{traefik} & 58.95\% & 59.13\% & 58.99\% & 59.93\% & 61.39\% \\
        \midrule
        \textbf{Average} & 56.49\% & 59.97\% & 58.24\% & 59.94\% & 60.98\% \\
        \bottomrule
        \end{tabular}%
    }
    \label{tab:rq1_enhance_the_original_unit_test}%
\end{table}%

\textbf{\underline{Effectiveness of \toolname in Mutation Testing.}}
Table~\ref{tab:rq1_mutation_testing} presents the results of the study.
For each project listed in column 1, the table details the number of generated mutants (column 2), the number of killed mutants by each approach (columns 3-6), and the mutator coverage percentages (columns 7-10).
As shown in Table~\ref{tab:rq1_mutation_testing}, we find that the unit tests generated by \toolname not only kill the highest number of mutants but also achieve the best mutator coverage across all evaluated projects.
For instance, in the ``gin'' project, a total of 885 mutants are generated.
NxtUnit, UniTester, and ChatUniTest kill 127, 61, and 153 mutants, respectively. 
In contrast, \toolname achieves an impressive 164 mutants killed, significantly surpassing the performance of the baselines.
Furthermore, \toolname achieves a mutator coverage of 26.05\%, which is notably higher than NxtUnit’s 18.21\%, UniTester's 12.73\%, and ChatUniTest’s 25.76\%.
These results underscore the effectiveness of \toolname in not only detecting defects but also improving the overall quality and reliability of the generated unit tests when compared to existing approaches.

\begin{table}[htbp]
    \centering
    \caption{RQ-1: \toolname vs. Baselines across different projects in mutation testing}
    \resizebox{\linewidth}{!}{
        \begin{tabular}{l|r|rrrr|rrrr}
        \toprule
        \multirow{1.5}[4]{*}{\textbf{Projects}} & \multirow{1.5}[4]{*}{\textbf{Mutants (\#)}} & \multicolumn{4}{c|}{\textbf{Killed Mutants (\#)}} & \multicolumn{4}{c}{\textbf{Mutator Coverage (\%)}} \\
        \cmidrule{3-10}      &   & \textbf{NxtUnit} & \textbf{UniTester} & \textbf{ChatUniTest} & \textbf{\toolname} & \textbf{NxtUnit} & \textbf{UniTester} & \textbf{ChatUniTest} & \textbf{\toolname} \\
        \midrule
        \textbf{beego} & 4,573  & 468 & 138 & 425 & 473 & 17.93\% & 8.19\% & 18.43\% & 19.91\% \\
        \textbf{echo} & 922  & 49 & 26 & 90 & 91 & 15.59\% & 12.21\% & 16.39\% & 16.59\% \\
        \textbf{fiber} & 1,554  & 153 & 80 & 141 & 160 & 15.24\% & 9.36\% & 15.93\% & 17.99\% \\
        \textbf{frp} & 1,538  & 84 & 57 & 83 & 99 & 6.95\% & 2.68\% & 6.88\% & 7.76\% \\
        \textbf{gin} & 885  & 127 & 61 & 153 & 164 & 18.21\% & 12.73\% & 25.76\% & 26.05\% \\
        \textbf{hugo} & 5,882  & 542 & 339 & 620 & 670 & 9.27\% & 7.75\% & 11.07\% & 11.36\% \\
        \textbf{nps} & 1,369  & 52 & 41 & 50 & 56 & 7.68\% & 4.29\% & 7.56\% & 8.36\% \\
        \textbf{traefik} & 4,331  & 269 & 109 & 264 & 308 & 7.35\% & 5.88\% & 7.67\% & 8.84\% \\
        \midrule
        \textbf{Average} & 2,632  & 218  & 106  & 228  & 253  & 12.28\% & 7.89\% & 13.71\% & 14.61\% \\
        \bottomrule
        \end{tabular}%
    }
    \label{tab:rq1_mutation_testing}%
\end{table}%

\intuition{
{\bf Answer to RQ-1}: 
\toolname significantly outperforms baselines in enhancing compile rate and line coverage, improving from 16.67\%–63.56\% to 45.58\%–69.49\% and from 7.49\%–53.92\% to 12.92\%–58.09\%, respectively.
It also surpasses other approaches in mutation testing, demonstrating its effectiveness in boosting software testing and quality.
}

\subsection{RQ-2: Model-Agnostic Capabilities of \toolname}
\label{sec:rq2}
\noindent 
\textbf{Objective.} 
In RQ-1, we use DeepSeek-Coder as the backbone model to evaluate the effectiveness of \toolname compared to the baselines. 
The results demonstrate that \toolname outperforms existing approaches and shows promising performance in generating unit test cases. 
In this RQ, we extend our analysis to examine the model-agnostic effectiveness of \toolname, specifically assessing whether \toolname maintains its effectiveness when applied to different models.

\noindent 
\textbf{Experimental Design.}
In addition to DeepSeek-Coder, we utilize two state-of-the-art open-source LLMs to investigate whether \toolname remains effective across different models, thus evaluating its model-agnostic capabilities. 
Specifically, the additional models are (1) CodeLlama~\cite{roziere2023code} and (2) Magicoder~\cite{wei2023magicoder}.
We follow the same experimental setup outlined in Section~\ref{sec:setup} and compare the performance of each model in terms of compile rate, line coverage, and mutation testing. 
To ensure consistency, we maintain identical experimental conditions across all basic LLMs and their corresponding \toolname implementations.

\noindent 
\textbf{Results.} We discuss the model-agnostic capabilities of \toolname from the aspects of compile rate, line coverage, and mutation testing, respectively.

\begin{figure}[!htbp]
\centerline{\includegraphics[width=\linewidth]{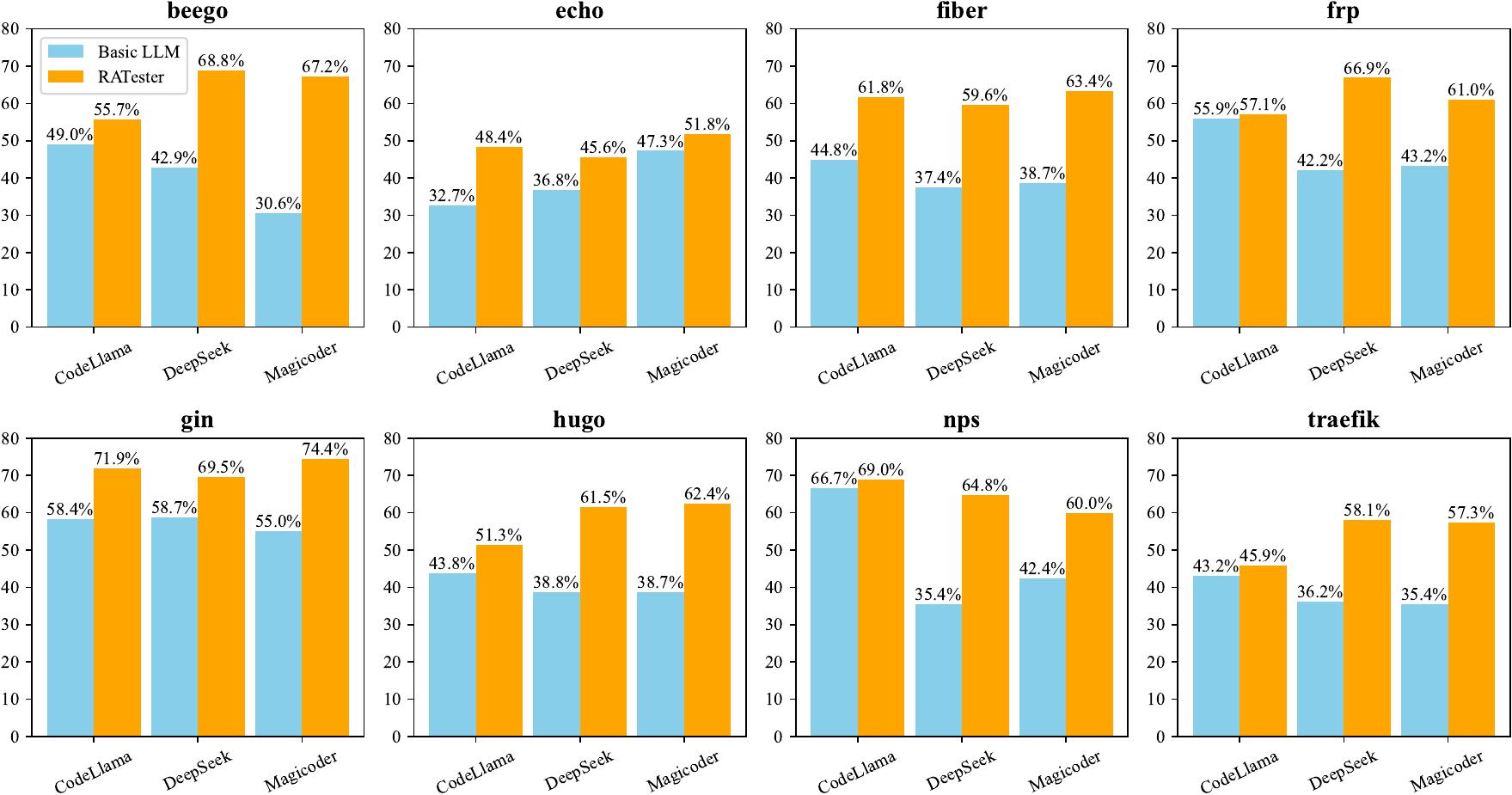}}
\caption{RQ-2: \toolname vs. Basic LLMs across different projects on compile rate}
\label{fig:rq2_compile_rate}
\end{figure}

\underline{\textbf{Model-agnostic capabilities of \toolname in compile rate.}}
Fig.~\ref{fig:rq2_compile_rate} shows the compile rate of unit tests generated by \toolname compared to basic LLMs. 
We find that \toolname significantly improves the performance of these basic LLMs. 
For instance, in the fiber project, \toolname increases the compile rate from 44.8\% to 61.8\% for CodeLlama, from 37.4\% to 59.6\% for DeepSeek-Coder, and from 38.7\% to 63.4\% for Magicoder. 
Overall, \toolname raises the compile rate of basic LLMs from a range of 30.6\%-66.7\% to 45.6\%-74.4\%, making more unit tests usable by developers during testing.
This not only demonstrates the effectiveness of the \toolname approach but also highlights its universal applicability. 
It is designed to be model-agnostic, meaning it can adapt to various LLMs, further emphasizing its flexibility and universality.

\begin{figure}[!htbp]
    \centering    
    \includegraphics[width=\linewidth]{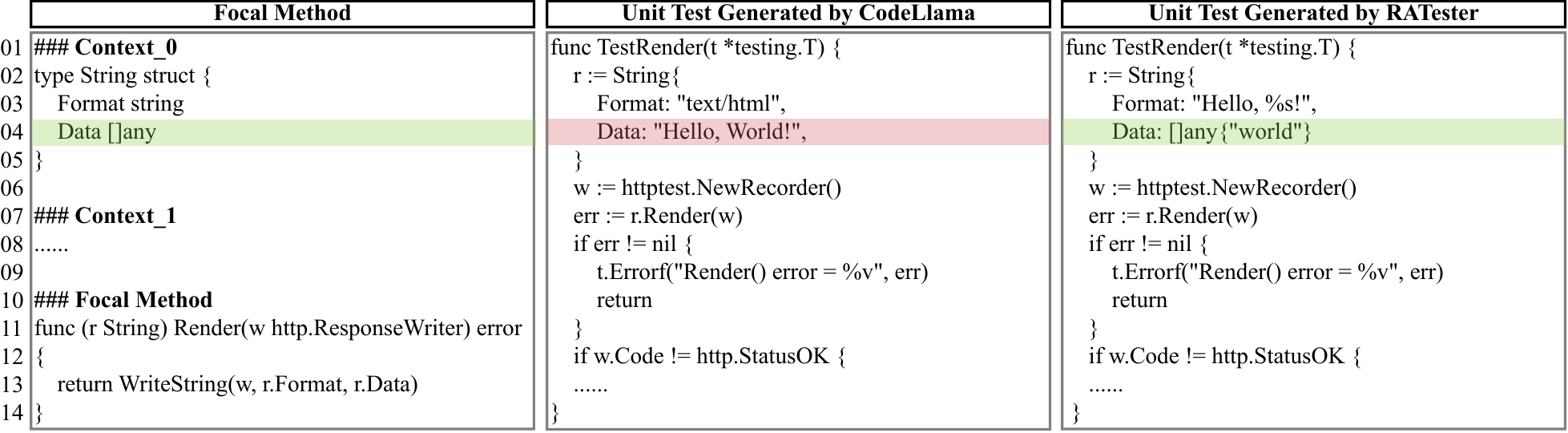}
    \caption{RQ-2: The unit tests generated by CodeLlama and \toolname for the gin project}
    \label{fig:unique}
\end{figure}

To further understand why \toolname has an outstanding performance in unit test generation, we conduct a case study by analyzing one example (i.e., the unit tests generated by CodeLlama and \toolname for the gin project) shown in Fig.~\ref{fig:unique}. 
On the left of Fig.~\ref{fig:unique} is the focal method, the middle presents the unit test generated by CodeLlama, and the right shows the one generated by \toolname.
For the test case generated by CodeLlama, it tests the ``Render'' method of a struct named ``String''. 
The test creates an instance of the ``String'' type, sets the content type to ``text/html'' and the data to ``Hello, World!''.
It then creates a mock HTTP response recorder using ``httptest.NewRecorder'', invokes the ``Render'' method to render the data, and checks if any errors occurred, if the HTTP status code is 200 OK, and if the response body correctly contains ``Hello, World!''. 
If an issue is detected at any checkpoint, the test reports an error using ``t.Errorf'' and terminates.
However, this test case contains a compilation error, with the following runtime error: ``cannot use `Hello, World!' (untyped string constant) as []any value in struct literal''.
This occurs because CodeLlama hallucinates when generating the test case for the focal method, lacking knowledge about the ``String'' struct (i.e., no code context). 
As a result, it incorrectly infers the ``Data'' field and mistakenly sets it as a string.
In contrast, \toolname dynamically retrieves the required definitions and documentation when generating test cases. 
As a result, it correctly identifies that the Data field in the ``String'' struct should be a slice of any type, not a single string. 
This allows \toolname to generate compilable unit test.

\underline{\textbf{Model-agnostic capabilities of \toolname in line coverage.}}
The left side of Table~\ref{tab:rq2_line_coverage} shows the line coverage results of basic LLMs, while the right side presents the results of \toolname using different backbone models.
We calculate not only the line coverage for \toolname with different backbone models but also the improvement percentages relative to basic LLMs.
As shown in the table, the \toolname approach increases the overall line coverage of unit tests generated by basic LLMs across different projects. For example, in the beego project, \toolname improves CodeLlama's coverage by 24.3\% (from 20.95\% to 26.04\%), DeepSeek-Coder's coverage by 40.3\% (from 22.75\% to 31.91\%), and Magicoder's coverage by 55.3\% (from 22.35\% to 34.71\%).
Overall, \toolname enhances the performance of basic LLMs, raising their average line coverage from a range of 18.80\%-19.95\% to 23.00\%-26.25\%, with relative improvements ranging from 22.3\% to 34.1\%. 
This aligns with the motivation behind our method's design: providing LLMs with more effective context (such as definitions of called methods) enhances the model's ability to generate unit tests and subsequently increases overall line coverage.

\begin{table}[htbp]
    \centering
    \caption{RQ-2: \toolname vs. Basic LLMs across different projects on line coverage}
    \resizebox{.9\linewidth}{!}{
        \begin{tabular}{l|ccc|ccc}
        \toprule
        \multirow{1.5}[4]{*}{\textbf{Projects}} & \multicolumn{3}{c|}{\textbf{Basic LLM}} & \multicolumn{3}{c}{\textbf{\toolname}} \\
        \cmidrule{2-7}      & \textbf{CodeLlama} & \textbf{DeepSeek} & \textbf{Magicoder} & \textbf{CodeLlama} & \textbf{DeepSeek} & \textbf{Magicoder} \\
        \midrule
        \textbf{beego} & 20.95\% & 22.75\% & 22.35\% &  26.04\% (↑24.3\%)  &  31.91\% (↑40.3\%)  &  34.71\% (↑55.3\%)  \\
        \textbf{echo} & 17.65\% & 21.08\% & 24.21\% &  26.35\% (↑49.3\%)  &  24.50\% (↑16.2\%)  &  27.55\% (↑13.8\%)  \\
        \textbf{fiber} & 14.94\% & 17.64\% & 14.82\% &  20.72\% (↑38.7\%)  &  26.31\% (↑49.1\%)  &  16.41\% (↑10.7\%)  \\
        \textbf{frp} & 11.68\% & 11.00\% & 12.95\% &  14.11\% (↑20.8\%)  &  14.43\% (↑31.2\%)  &  18.54\% (↑43.2\%)  \\
        \textbf{gin} & 43.53\% & 45.35\% & 42.67\% &  48.23\% (↑10.8\%)  &  58.09\% (↑28.1\%)  &  47.13\% (↑10.5\%)  \\
        \textbf{hugo} & 16.10\% & 16.87\% & 16.02\% &  19.77\% (↑22.8\%)  &  25.02\% (↑48.3\%)  &  18.92\% (↑18.1\%)  \\
        \textbf{nps} & 11.83\% & 10.22\% & 11.33\% &  13.12\% (↑10.9\%)  &  16.82\% (↑64.6\%)  &  14.93\% (↑31.8\%)  \\
        \textbf{traefik} & 13.72\% & 11.68\% & 15.26\% &  15.64\% (↑14.0\%)  &  12.92\% (↑10.6\%)  &  18.62\% (↑22.0\%)  \\
        \midrule
        \textbf{Average} & 18.80\% & 19.57\% & 19.95\% &  23.00\% (↑22.3\%)  &  26.25\% (↑34.1\%)  &  24.60\% (↑23.3\%)  \\
        \bottomrule
        \end{tabular}%
    }
    \label{tab:rq2_line_coverage}%
\end{table}%

\underline{\textbf{Model-agnostic capabilities of \toolname in mutation testing.}}
Table~\ref{tab:rq2_mutation_testing} presents the results of \toolname in using different backbone models in terms of mutation testing. 
From the results, we find that:
\textbf{(1) Performance Variation Across Backbone Models:} There are significant performance differences among the basic LLMs, which directly impacts the effectiveness of \toolname. 
Among the various backbone models tested, DeepSeek-Coder demonstrates superior performance, leading to the highest effectiveness when \toolname utilizes DeepSeek-Coder as its backbone model.
\textbf{(2) Enhancement Across All Models:} \toolname consistently improves the performance of all three basic LLMs utilized in this study. 
For instance, in the hugo project, Magicoder kills only 522 mutants.
In contrast, \toolname using Magicoder successfully kills 583 mutants, showcasing a clear enhancement in defect detection capabilities.
\textbf{(3) Overall Improvement in Defect Detection:} Across all three models tested, \toolname exhibits a significant advantage, killing between 316 and 522 more mutants compared to the basic LLMs. 
This indicates that \toolname not only leverages the strengths of the underlying models but also enhances their overall effectiveness in detecting defects.

\begin{table}[htbp]
    \centering
    \caption{RQ-2: \toolname vs. Basic LLMs across different projects in mutation testing}
    \resizebox{\linewidth}{!}{
        \begin{tabular}{l|r|rrr|rrr}
        \toprule
        \multirow{1.5}[4]{*}{\textbf{Projects}} & \multirow{1.5}[4]{*}{\textbf{Mutants (\#)}} & \multicolumn{3}{c|}{\textbf{Basic LLM}} & \multicolumn{3}{c}{\textbf{\toolname}} \\
        \cmidrule{3-8}      &   & \textbf{CodeLlama} & \textbf{DeepSeek} & \textbf{Magicoder} & \textbf{CodeLlama} & \textbf{DeepSeek} & \textbf{Magicoder} \\
        \midrule
        \textbf{beego} & 4,573  & 209  & 381  & 224  & 388 (+179) & 473 (+92) & 524 (+300) \\
        \textbf{echo} & 922  & 42  & 48  & 50  & 48 (+6) & 91 (+43) & 61 (+11) \\
        \textbf{fiber} & 1,554  & 126  & 131  & 115  & 131 (+5) & 160 (+29) & 124 (+9) \\
        \textbf{frp} & 1,538  & 67  & 70  & 69  & 80 (+13) & 99 (+29) & 113 (+44) \\
        \textbf{gin} & 885  & 59  & 135  & 86  & 78 (+19) & 164 (+29) & 115 (+29) \\
        \textbf{hugo} & 5,882  & 428  & 615  & 522  & 448 (+20) & 670 (+55) & 583 (+61) \\
        \textbf{nps} & 1,369  & 34  & 50  & 46  & 46 (+12) & 56 (+6) & 56 (+10) \\
        \textbf{traefik} & 4,331  & 145  & 241  & 156  & 207 (+62) & 308 (+67) & 214 (+58) \\
        \midrule
        \textbf{Sum} & 21,054  & 1,110  & 1,671  & 1,268  & 1,426 (+316) & 2,021 (+350) & 1,790 (+522) \\
        \bottomrule
        \end{tabular}%
    }
    \label{tab:rq2_mutation_testing}%
\end{table}%

\intuition{
{\bf Answer to RQ-2}: 
The basic LLMs have limited capabilities in generating unit tests, while \toolname enhances these capabilities through appropriate adaptations. 
Overall, \toolname significantly outperforms the basic LLMs in compile rate, line coverage, and mutation testing, demonstrating its adaptability and improved effectiveness.
}

\subsection{RQ-3: Efficiency of \toolname}
\label{sec:rq3}
\noindent 
\textbf{Objective.} 
The previous research question examined the effectiveness of \toolname. 
In this RQ, we aim to study the efficiency of \toolname.
We conduct a comprehensive experiment to evaluate its efficiency, focusing not only on the time required to generate test cases but also on the impact of the candidate number of generated unit tests for each focal method or function.

\noindent 
\textbf{Experimental Design.}
We begin by investigating the total time required to generate test cases for all eight projects.
We use the baselines (i.e., NxtUnit, UniTester, ChatUniTest, CodeLlama, DeepSeek-Coder, and Magicoder) to compare the total generation time of \toolname across all projects.
Additionally, we conduct a comparative analysis of the number of unit test candidates. 
We use DeepSeek-Coder and \toolname (with DeepSeek-Coder as the backbone model) to generate test cases for all projects, setting the number of unit test candidates from 1 to 10 (i.e., generating 1 to 10 test cases for each focal method or function). 
We then calculate the average line coverage across all projects.

\noindent 
\textbf{Results.} We discuss the results from two perspectives: the total generation time across all projects and the impact of the unit test candidate number.

\begin{figure}[!htbp]
    \centering
    \includegraphics[width=\linewidth]{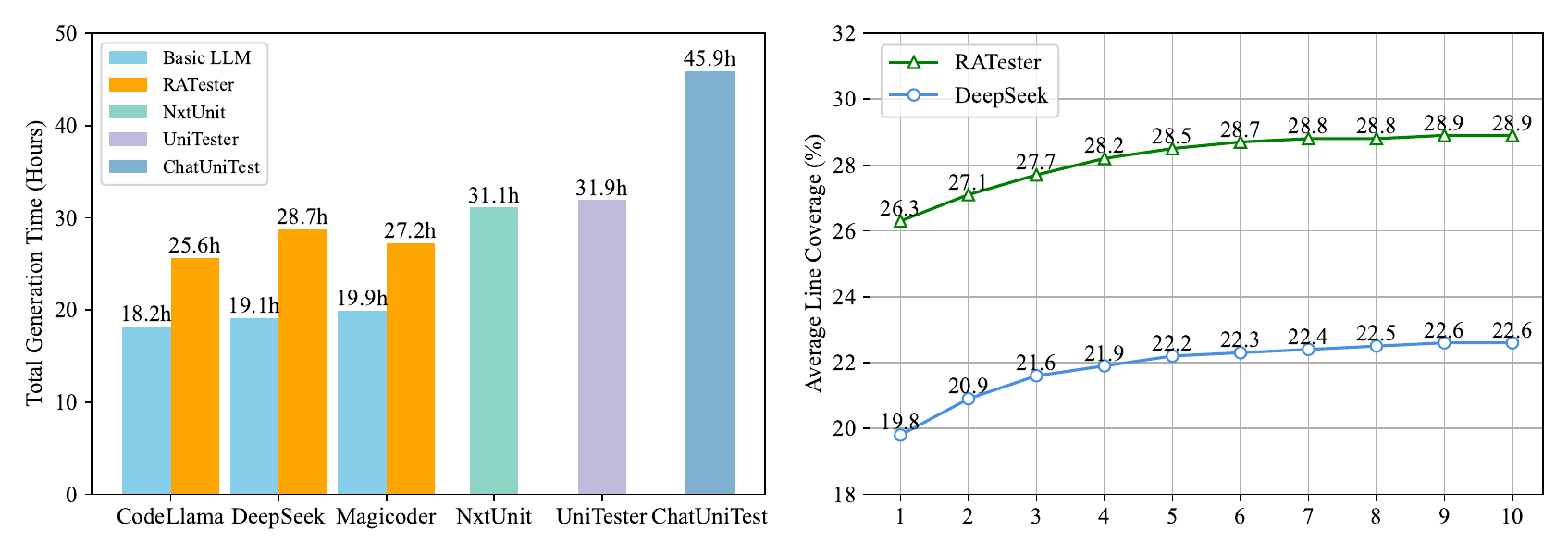}
    \caption{RQ-3: (Left) Total generation time of \toolname and baselines; (Right) Performance of \toolname and DeepSeek-Coder across varying unit test candidate number}
    \label{fig:rq3}
\end{figure}

\underline{\textbf{Total generation time across all projects.}}
The left side of Fig.~\ref{fig:rq3} shows the total generation time required by each baseline and \toolname for unit test generation. 
According to the results, we observe that:
(1) \toolname requires more generation time compared to the basic LLMs.
For example, CodeLlama alone takes 18.2 hours, while \toolname built on CodeLlama requires 25.6 hours. 
Overall, the basic LLMs need between 18.2 and 19.9 hours, whereas \toolname requires between 25.6 and 28.7 hours. 
However, we believe this additional time is acceptable in practical usage due to the higher compile rates, increased line coverage, and a greater number of killed mutants achieved by \toolname (refer to Section~\ref{sec:rq2} for more details).
(2) The three basic LLMs (i.e., CodeLlama, DeepSeek-Coder, and Magicoder) perform fast, while the others perform relatively a little slow.
More precisely, CodeLlama, DeepSeek-Coder, and Magicoder only take 18.2 hours, 19.1 hours, and 19.9 hours to generate unit tests for all projects.
Among all the approaches, ChatUniTest has the longest generation time, taking 45.9 hours.
This is because ChatUniTest includes redundant context and also features a unit test repair process that requires multiple iterations to arrive at the final unit test.

\underline{\textbf{Impact of the unit test candidate number.}}
According to the results on the right side of Fig.~\ref{fig:rq3}, we find that:
(1) Different candidate numbers have varying impacts on the performance of \toolname and DeepSeek-Coder, with both models showing improved performance as the number of candidates increases.
(2) The curve indicates that the performance of \toolname far exceeds that of DeepSeek-Coder.
\toolname only requires generating one unit test for each focal method or function to achieve higher line coverage than DeepSeek-Coder, which requires ten unit tests. 
This demonstrates that \toolname is more efficient and produces higher-quality unit tests with a less number of candidates.
(3) Increasing the number of candidates does not guarantee significant performance improvements.
As we continuously increase the number of candidates from 2 to 10, the performance of both \toolname and DeepSeek-Coder improves only slightly, while the generation cost with the LLM increases significantly.
Considering both the performance improvements and the associated generation costs, we adopt a candidate number of 1 unit test as the default setting.

\intuition{
{\bf Answer to RQ-3}: 
(1) \toolname requires more generation time than the basic LLMs; however, considering the performance improvements, this additional time is justified.
(2) Increasing the candidate number can enhance the performance of \toolname, but the improvement is not significant.
}
\label{sec:results}

\section{Threats to Validity}
\noindent
\textbf{Internal Validity.}
The first one arises from potential data leakage since referenced unit tests may be part of the training data of LLMs (e.g., CodeLlama and DeepSeek-Coder).
To tackle this issue, we initially calculate the number of unit tests generated by \toolname, which matches the reference unit test in all eight projects.
We find that out of 11,195 generated unit tests, only 1 of these aligns with the unit tests in projects. 
Additionally, compared to the basic LLMs (i.e., CodeLlama, DeepSeek-Coder, and Magicoder), \toolname demonstrates a significant enhancement in performance, achieving an increase in line coverage of 22.3\% to 34.1\%.
Furthermore, the unit tests generated by \toolname also contribute to completing the original unit tests in the projects, raising the line coverage from 56.49\% to 60.98\%.
This demonstrates that the improved results achieved by \toolname are not merely a result of memorizing the training data.

The second concern arises from potential errors in implementing our approach and baselines.
To mitigate this threat, we implement our model through pair programming, using the source code of baselines provided by the corresponding authors and adhering to the same settings outlined in the original papers. 
Additionally, the authors conduct a thorough review of the experimental scripts to ensure their correctness.

\noindent
\textbf{External Validity.}
The main external threat to validity comes from our evaluation dataset used. 
The effectiveness demonstrated by \toolname may not be generalizable to different unit test generation datasets, particularly those involving unit tests written in other programming languages (e.g., Java and Python).
This limitation is common to some pipelines that utilize LLMs and language-specific tools (i.e., gopls), and we aim to address it in our future work.
\label{sec:threats}

\section{Related Work}
\subsection{Unit Test Generation}

Unit test generation approaches can be classified into three types: traditional approaches, learning-based approaches, and LLM-based approaches.

Traditional approaches~\cite{fraser2011evosuite, pacheco2007randoop} focus on code coverage, and research shows that traditional approaches are very effective at achieving high coverage~\cite{aleti2017analysing, oliveira2018mapping, panichella2015reformulating, panichella2017automated}.
Pacheco et al. introduced Randoop~\cite{pacheco2007randoop}, a widely adopted tool that generates unit tests for Java.
Fraser et al. developed EvoSuite~\cite{fraser2011evosuite}, which automatically creates test suites optimized for high coverage, minimal size, and rich assertions.
However, previous studies show that these approaches do not produce well-written, maintainable unit tests explicitly for developers to use~\cite{wang2023nxtunit, tufano2020unit}.

To address limitations in traditional unit test generation, learning-based approaches~\cite{alagarsamy2023a3test, tufano2020unit, dinella2022toga, saes2018unit, he2024unitsyn} have made notable advancements. 
Saes~\cite{saes2018unit} leveraged over 780K focal-test method pairs from GitHub’s JUnit framework to generate Java test suites with an 86.69\% parsability rate.
Tufano et al.~\cite{tufano2020unit} introduced AthenaTest, which fine-tunes the BART model on the Methods2Test dataset~\cite{tufano2022methods2test} to generate entire unit tests from focal method contexts, achieving a correct test rate for 43\% of focal methods, with 16\% classified as valid tests.
Dinella et al.~\cite{dinella2022toga} proposed TOGA, which is a unified transformer-based neural approach to infer both exceptional and assertion test oracles based on the context of the focal method.
Alagarsamy et al.~\cite{alagarsamy2023a3test} developed A3Test, which combines test oracle generation with domain adaptation to enhance naming consistency and signature verification, outperforming AthenaTest in test accuracy and method coverage. 
He et al.~\cite{he2024unitsyn} proposed UniTester, which is trained on the UniTSyn dataset and is capable of synthesizing unit tests for programs in multiple languages, including Golang.
However, these approaches rely heavily on task-specific datasets extracted from open-source repositories.

In response to the challenges posed by learning-based approaches, researchers are increasingly utilizing pre-trained LLMs to generate unit tests directly from contextual information, reducing reliance on task-specific datasets by leveraging extensive pre-training on diverse open-source code.
Lemieux et al.~\cite{lemieux2023codamosa} introduced CodaMOSA, an SBST approach that leverages LLMs to overcome coverage plateaus in Python code. 
Following that, Ni et al.~\cite{ni2024casmodatest} also explored ChatGPT for generating unit tests based on focal methods.
Despite these advancements, LLMs may still exhibit hallucinations when generating unit tests for focal methods and functions due to their lack of the project’s global knowledge.
These hallucinations can include but are not limited to, calling non-existent methods, as well as assigning incorrect parameters and return values (e.g., mismatched parameter types or incorrect parameter counts).
To overcome this limitation, many studies have explored the extraction of context to reduce hallucinations in the generation process of LLMs. 
Yuan et al.~\cite{yuan2023no} developed ChatTester, an LLM-based model employing ChatGPT with an iterative generate-and-validate strategy that incorporates execution feedback and code context. 
Chen et al.~\cite{chen2024chatunitest} proposed ChatUniTest, which is an innovative framework designed to enhance automated unit test generation. 
It utilizes an LLM-based approach, augmented with an adaptive focal context mechanism to capture relevant context in prompts, and employs a ``Generation-Validation-Repair'' process to correct errors in generated tests. 
Following that, researchers~\cite{ryan2024code, gao2025prompt,yuan2024evaluating} have explored the roles of focal context and dependency context.
These methods utilize one or more fixed patterns to extract context for the focal method: (1) focal class signature; (2) signatures of other methods and fields in the class; (3) signatures of dependent classes; and (4) signatures of dependent methods and fields in the dependent classes.
However, these fixed extraction patterns present several issues: (1) they may overlook important context; for instance, when generating a unit test for a specific focal method, the LLM might require unknown context beyond the dependencies of that focal method; (2) there is potential for redundant context, as excessive irrelevant context could lead to redundancy, preventing the model from focusing on essential information.

Different from existing works, our paper presents an LLM-based framework that not only eliminates the reliance on task-specific datasets but also proactively utilizes gopls to fetch precise context. 
This approach significantly reduces hallucinations during the test generation process.

\subsection{Pre-trained Language Model}
With advancements in Natural Language Processing, Pre-trained Language Models have gained widespread traction due to their ability to be trained on billions of parameters and vast datasets, which has led to remarkable performance improvements across diverse applications. 
These models are highly adaptable to a range of downstream tasks, utilizing methods such as fine-tuning~\cite{radford2018improving,yin2024multitask} and prompting~\cite{liu2023pre,yin2024thinkrepair,yin2024rectifier,ni2024casmodatest}. 
Their versatility arises from extensive pre-training on broad data, equipping them with a robust knowledge base applicable across numerous domains.
Fine-tuning involves adjusting model parameters specifically for a targeted task, requiring iterative training on a dedicated dataset, which enhances the model's accuracy and relevance for that task.
By contrast, prompting offers a more direct, efficient approach by feeding the model task-specific instructions or a few relevant examples in natural language, allowing it to perform effectively without parameter adjustments.
Although fine-tuning can yield higher accuracy, it demands significant computational resources and is less feasible in scenarios with limited task-specific datasets.

Pre-trained Language Models are typically based on the transformer architecture~\cite{vaswani2017attention} and are categorized into three types: encoder-only, encoder-decoder, and decoder-only architectures.
Encoder-only models, such as 
CodeBERT~\cite{feng2020codebert} and GraphCodeBERT~\cite{guo2020graphcodebert}, and encoder-decoder models, like PLBART~\cite{ahmad2021unified} and CodeT5~\cite{wang2021codet5}, are trained using objectives like Masked Language Modeling (MLM) or Masked Span Prediction (MSP).
In these setups, a small percentage (e.g., 15\%) of the tokens are replaced with either masked tokens or masked span tokens, and the model learns to predict or recover the masked content. 
Trained on diverse code-related data, these models are then fine-tuned for specific tasks to achieve enhanced performance~\cite{ni2023distinguishing,fu2022linevul,hin2022linevd}.
Decoder-only models have gained significant attention, primarily due to their use of causal language modeling objectives, which train them to predict the probability of each next token based on all previous tokens in a sequence. GPT~\cite{radford2018improving} and its variants are the most prominent examples of this architecture, marking a pivotal point in bringing large language models into widespread practical applications.

To enhance LLMs' generalization and alignment with human intentions on previously unseen downstream tasks, recent research has focused on instruction tuning and reinforcement learning to improve model performance~\cite{christiano2017deep,ouyang2022training,ziegler2019fine}. 
For instance, OpenAI’s ChatGPT~\cite{openai2022chatgpt} is a notable example built on the generative pre-trained transformer architecture. 
It undergoes initial instruction tuning, followed by updates through reinforcement learning from human feedback to better capture human-aligned responses.
Beyond commercial models, there is also a growing landscape of open-source instructed LLMs, such as CodeLlama~\cite{roziere2023code} and DeepSeek-Coder~\cite{deepseek-coder}, which demonstrate promising performance across various tasks and hold potential for broader adaptability~\cite{yin2024multitask,ni2024casmodatest,yin2024you}.
\label{sec:related_work}

\section{Conclusion}
This paper enhances the LLM’s ability to generate more repository-aware unit tests through global contextual information injection.
To provide LLMs with a level of global knowledge similar to that of human testers, \toolname integrates the language server gopls.
When it encounters unfamiliar identifiers, such as struct names, \toolname utilizes gopls to fetch relevant precise context, thereby preventing erroneous usage.
This integration enriches the LLM’s global knowledge of the project, significantly reducing hallucinations.
We evaluate the effectiveness and efficiency of \toolname by constructing a new Golang dataset from real-world projects and comparing it against baseline approaches. 
The results illustrate the advantages of \toolname over these baselines.
Furthermore, we extend our analysis to assess the model-agnostic effectiveness of \toolname. 
These findings not only validate the efficacy of \toolname but also emphasize its universal applicability.
\label{sec:conclusion}


\section*{Acknowledgements}{
This work was supported by the National Natural Science Foundation of China (Grant No.62202419), the Fundamental Research Funds for the Central Universities (No. 226-2022-00064),
Zhejiang Provincial Natural Science Foundation of China (No. LY24F020008),
the Ningbo Natural Science Foundation (No. 2022J184), 
and the State Street Zhejiang University Technology Center.
}

\balance
\bibliographystyle{ACM-Reference-Format}
\bibliography{main}


\begin{thebibliography}{55}


\ifx \showCODEN    \undefined \def \showCODEN     #1{\unskip}     \fi
\ifx \showDOI      \undefined \def \showDOI       #1{#1}\fi
\ifx \showISBNx    \undefined \def \showISBNx     #1{\unskip}     \fi
\ifx \showISBNxiii \undefined \def \showISBNxiii  #1{\unskip}     \fi
\ifx \showISSN     \undefined \def \showISSN      #1{\unskip}     \fi
\ifx \showLCCN     \undefined \def \showLCCN      #1{\unskip}     \fi
\ifx \shownote     \undefined \def \shownote      #1{#1}          \fi
\ifx \showarticletitle \undefined \def \showarticletitle #1{#1}   \fi
\ifx \showURL      \undefined \def \showURL       {\relax}        \fi
\providecommand\bibfield[2]{#2}
\providecommand\bibinfo[2]{#2}
\providecommand\natexlab[1]{#1}
\providecommand\showeprint[2][]{arXiv:#2}

\bibitem[gop(2024)]%
        {gopls}
 \bibinfo{year}{2024}\natexlab{}.
\newblock \bibinfo{title}{gopls}.
\newblock
\newblock
\urldef\tempurl%
\url{https://github.com/golang/tools/tree/master/gopls}
\showURL{%
\tempurl}


\bibitem[gre(2024)]%
        {gremlins}
 \bibinfo{year}{2024}\natexlab{}.
\newblock \bibinfo{title}{gremlins}.
\newblock
\newblock
\urldef\tempurl%
\url{https://github.com/go-gremlins/gremlins}
\showURL{%
\tempurl}


\bibitem[hug(2024)]%
        {huggingface}
 \bibinfo{year}{2024}\natexlab{}.
\newblock \bibinfo{title}{Hugging Face}.
\newblock
\newblock
\urldef\tempurl%
\url{https://huggingface.co}
\showURL{%
\tempurl}


\bibitem[Ahmad et~al\mbox{.}(2021)]%
        {ahmad2021unified}
\bibfield{author}{\bibinfo{person}{Wasi~Uddin Ahmad}, \bibinfo{person}{Saikat Chakraborty}, \bibinfo{person}{Baishakhi Ray}, {and} \bibinfo{person}{Kai-Wei Chang}.} \bibinfo{year}{2021}\natexlab{}.
\newblock \showarticletitle{Unified pre-training for program understanding and generation}.
\newblock \bibinfo{journal}{\emph{arXiv preprint arXiv:2103.06333}} (\bibinfo{year}{2021}).
\newblock


\bibitem[AI(2023)]%
        {deepseek-coder}
\bibfield{author}{\bibinfo{person}{DeepSeek AI}.} \bibinfo{year}{2023}\natexlab{}.
\newblock \bibinfo{title}{DeepSeek Coder: Let the Code Write Itself}.
\newblock \bibinfo{howpublished}{\url{https://github.com/deepseek-ai/DeepSeek-Coder}}.
\newblock


\bibitem[Alagarsamy et~al\mbox{.}(2023)]%
        {alagarsamy2023a3test}
\bibfield{author}{\bibinfo{person}{Saranya Alagarsamy}, \bibinfo{person}{Chakkrit Tantithamthavorn}, {and} \bibinfo{person}{Aldeida Aleti}.} \bibinfo{year}{2023}\natexlab{}.
\newblock \showarticletitle{A3Test: Assertion-Augmented Automated Test Case Generation}.
\newblock \bibinfo{journal}{\emph{arXiv preprint arXiv:2302.10352}} (\bibinfo{year}{2023}).
\newblock


\bibitem[Aleti et~al\mbox{.}(2017)]%
        {aleti2017analysing}
\bibfield{author}{\bibinfo{person}{Aldeida Aleti}, \bibinfo{person}{Irene Moser}, {and} \bibinfo{person}{Lars Grunske}.} \bibinfo{year}{2017}\natexlab{}.
\newblock \showarticletitle{Analysing the fitness landscape of search-based software testing problems}.
\newblock \bibinfo{journal}{\emph{Automated Software Engineering}}  \bibinfo{volume}{24} (\bibinfo{year}{2017}), \bibinfo{pages}{603--621}.
\newblock


\bibitem[Barr et~al\mbox{.}(2014)]%
        {barr2014oracle}
\bibfield{author}{\bibinfo{person}{Earl~T Barr}, \bibinfo{person}{Mark Harman}, \bibinfo{person}{Phil McMinn}, \bibinfo{person}{Muzammil Shahbaz}, {and} \bibinfo{person}{Shin Yoo}.} \bibinfo{year}{2014}\natexlab{}.
\newblock \showarticletitle{The oracle problem in software testing: A survey}.
\newblock \bibinfo{journal}{\emph{IEEE transactions on software engineering}} \bibinfo{volume}{41}, \bibinfo{number}{5} (\bibinfo{year}{2014}), \bibinfo{pages}{507--525}.
\newblock


\bibitem[Chen et~al\mbox{.}(2024)]%
        {chen2024chatunitest}
\bibfield{author}{\bibinfo{person}{Yinghao Chen}, \bibinfo{person}{Zehao Hu}, \bibinfo{person}{Chen Zhi}, \bibinfo{person}{Junxiao Han}, \bibinfo{person}{Shuiguang Deng}, {and} \bibinfo{person}{Jianwei Yin}.} \bibinfo{year}{2024}\natexlab{}.
\newblock \showarticletitle{Chatunitest: A framework for llm-based test generation}. In \bibinfo{booktitle}{\emph{Companion Proceedings of the 32nd ACM International Conference on the Foundations of Software Engineering}}. \bibinfo{pages}{572--576}.
\newblock


\bibitem[Christiano et~al\mbox{.}(2017)]%
        {christiano2017deep}
\bibfield{author}{\bibinfo{person}{Paul~F Christiano}, \bibinfo{person}{Jan Leike}, \bibinfo{person}{Tom Brown}, \bibinfo{person}{Miljan Martic}, \bibinfo{person}{Shane Legg}, {and} \bibinfo{person}{Dario Amodei}.} \bibinfo{year}{2017}\natexlab{}.
\newblock \showarticletitle{Deep reinforcement learning from human preferences}.
\newblock \bibinfo{journal}{\emph{Advances in neural information processing systems}}  \bibinfo{volume}{30} (\bibinfo{year}{2017}).
\newblock


\bibitem[Dinella et~al\mbox{.}(2022)]%
        {dinella2022toga}
\bibfield{author}{\bibinfo{person}{Elizabeth Dinella}, \bibinfo{person}{Gabriel Ryan}, \bibinfo{person}{Todd Mytkowicz}, {and} \bibinfo{person}{Shuvendu~K Lahiri}.} \bibinfo{year}{2022}\natexlab{}.
\newblock \showarticletitle{Toga: A neural method for test oracle generation}. In \bibinfo{booktitle}{\emph{Proceedings of the 44th International Conference on Software Engineering}}. \bibinfo{pages}{2130--2141}.
\newblock


\bibitem[Feng et~al\mbox{.}(2020)]%
        {feng2020codebert}
\bibfield{author}{\bibinfo{person}{Zhangyin Feng}, \bibinfo{person}{Daya Guo}, \bibinfo{person}{Duyu Tang}, \bibinfo{person}{Nan Duan}, \bibinfo{person}{Xiaocheng Feng}, \bibinfo{person}{Ming Gong}, \bibinfo{person}{Linjun Shou}, \bibinfo{person}{Bing Qin}, \bibinfo{person}{Ting Liu}, \bibinfo{person}{Daxin Jiang}, {et~al\mbox{.}}} \bibinfo{year}{2020}\natexlab{}.
\newblock \showarticletitle{Codebert: A pre-trained model for programming and natural languages}.
\newblock \bibinfo{journal}{\emph{arXiv preprint arXiv:2002.08155}} (\bibinfo{year}{2020}), \bibinfo{pages}{1536--1547}.
\newblock


\bibitem[Fraser and Arcuri(2011)]%
        {fraser2011evosuite}
\bibfield{author}{\bibinfo{person}{Gordon Fraser} {and} \bibinfo{person}{Andrea Arcuri}.} \bibinfo{year}{2011}\natexlab{}.
\newblock \showarticletitle{Evosuite: automatic test suite generation for object-oriented software}. In \bibinfo{booktitle}{\emph{Proceedings of the 19th ACM SIGSOFT symposium and the 13th European conference on Foundations of software engineering}}. \bibinfo{pages}{416--419}.
\newblock


\bibitem[Fu and Tantithamthavorn(2022)]%
        {fu2022linevul}
\bibfield{author}{\bibinfo{person}{Michael Fu} {and} \bibinfo{person}{Chakkrit Tantithamthavorn}.} \bibinfo{year}{2022}\natexlab{}.
\newblock \showarticletitle{Linevul: A transformer-based line-level vulnerability prediction}. In \bibinfo{booktitle}{\emph{Proceedings of the 19th International Conference on Mining Software Repositories}}. \bibinfo{pages}{608--620}.
\newblock


\bibitem[Gao et~al\mbox{.}(2025)]%
        {gao2025prompt}
\bibfield{author}{\bibinfo{person}{Shuzheng Gao}, \bibinfo{person}{Chaozheng Wang}, \bibinfo{person}{Cuiyun Gao}, \bibinfo{person}{Xiaoqian Jiao}, \bibinfo{person}{Chun~Yong Chong}, \bibinfo{person}{Shan Gao}, {and} \bibinfo{person}{Michael Lyu}.} \bibinfo{year}{2025}\natexlab{}.
\newblock \showarticletitle{The Prompt Alchemist: Automated LLM-Tailored Prompt Optimization for Test Case Generation}.
\newblock \bibinfo{journal}{\emph{arXiv preprint arXiv:2501.01329}} (\bibinfo{year}{2025}).
\newblock


\bibitem[Garousi and Zhi(2013)]%
        {garousi2013survey}
\bibfield{author}{\bibinfo{person}{Vahid Garousi} {and} \bibinfo{person}{Junji Zhi}.} \bibinfo{year}{2013}\natexlab{}.
\newblock \showarticletitle{A survey of software testing practices in Canada}.
\newblock \bibinfo{journal}{\emph{Journal of Systems and Software}} \bibinfo{volume}{86}, \bibinfo{number}{5} (\bibinfo{year}{2013}), \bibinfo{pages}{1354--1376}.
\newblock


\bibitem[Guo et~al\mbox{.}(2020)]%
        {guo2020graphcodebert}
\bibfield{author}{\bibinfo{person}{Daya Guo}, \bibinfo{person}{Shuo Ren}, \bibinfo{person}{Shuai Lu}, \bibinfo{person}{Zhangyin Feng}, \bibinfo{person}{Duyu Tang}, \bibinfo{person}{Shujie Liu}, \bibinfo{person}{Long Zhou}, \bibinfo{person}{Nan Duan}, \bibinfo{person}{Alexey Svyatkovskiy}, \bibinfo{person}{Shengyu Fu}, {et~al\mbox{.}}} \bibinfo{year}{2020}\natexlab{}.
\newblock \showarticletitle{Graphcodebert: Pre-training code representations with data flow}.
\newblock \bibinfo{journal}{\emph{arXiv preprint arXiv:2009.08366}} (\bibinfo{year}{2020}).
\newblock


\bibitem[He et~al\mbox{.}(2024)]%
        {he2024unitsyn}
\bibfield{author}{\bibinfo{person}{Yifeng He}, \bibinfo{person}{Jiabo Huang}, \bibinfo{person}{Yuyang Rong}, \bibinfo{person}{Yiwen Guo}, \bibinfo{person}{Ethan Wang}, {and} \bibinfo{person}{Hao Chen}.} \bibinfo{year}{2024}\natexlab{}.
\newblock \showarticletitle{UniTSyn: A Large-Scale Dataset Capable of Enhancing the Prowess of Large Language Models for Program Testing}. In \bibinfo{booktitle}{\emph{Proceedings of the 33rd ACM SIGSOFT International Symposium on Software Testing and Analysis}}. \bibinfo{pages}{1061--1072}.
\newblock


\bibitem[Hin et~al\mbox{.}(2022)]%
        {hin2022linevd}
\bibfield{author}{\bibinfo{person}{David Hin}, \bibinfo{person}{Andrey Kan}, \bibinfo{person}{Huaming Chen}, {and} \bibinfo{person}{M~Ali Babar}.} \bibinfo{year}{2022}\natexlab{}.
\newblock \showarticletitle{LineVD: Statement-level Vulnerability Detection using Graph Neural Networks}.
\newblock \bibinfo{journal}{\emph{arXiv preprint arXiv:2203.05181}} (\bibinfo{year}{2022}).
\newblock


\bibitem[Lee et~al\mbox{.}(2012)]%
        {lee2012survey}
\bibfield{author}{\bibinfo{person}{Jihyun Lee}, \bibinfo{person}{Sungwon Kang}, {and} \bibinfo{person}{Danhyung Lee}.} \bibinfo{year}{2012}\natexlab{}.
\newblock \showarticletitle{Survey on software testing practices}.
\newblock \bibinfo{journal}{\emph{IET software}} \bibinfo{volume}{6}, \bibinfo{number}{3} (\bibinfo{year}{2012}), \bibinfo{pages}{275--282}.
\newblock


\bibitem[Lemieux et~al\mbox{.}(2023)]%
        {lemieux2023codamosa}
\bibfield{author}{\bibinfo{person}{Caroline Lemieux}, \bibinfo{person}{Jeevana~Priya Inala}, \bibinfo{person}{Shuvendu~K Lahiri}, {and} \bibinfo{person}{Siddhartha Sen}.} \bibinfo{year}{2023}\natexlab{}.
\newblock \showarticletitle{Codamosa: Escaping coverage plateaus in test generation with pre-trained large language models}. In \bibinfo{booktitle}{\emph{2023 IEEE/ACM 45th International Conference on Software Engineering (ICSE)}}. IEEE, \bibinfo{pages}{919--931}.
\newblock


\bibitem[Li et~al\mbox{.}(2023)]%
        {li2023starcoder}
\bibfield{author}{\bibinfo{person}{Raymond Li}, \bibinfo{person}{Loubna~Ben Allal}, \bibinfo{person}{Yangtian Zi}, \bibinfo{person}{Niklas Muennighoff}, \bibinfo{person}{Denis Kocetkov}, \bibinfo{person}{Chenghao Mou}, \bibinfo{person}{Marc Marone}, \bibinfo{person}{Christopher Akiki}, \bibinfo{person}{Jia Li}, \bibinfo{person}{Jenny Chim}, {et~al\mbox{.}}} \bibinfo{year}{2023}\natexlab{}.
\newblock \showarticletitle{StarCoder: may the source be with you!}
\newblock \bibinfo{journal}{\emph{arXiv preprint arXiv:2305.06161}} (\bibinfo{year}{2023}).
\newblock


\bibitem[Liu et~al\mbox{.}(2023)]%
        {liu2023pre}
\bibfield{author}{\bibinfo{person}{Pengfei Liu}, \bibinfo{person}{Weizhe Yuan}, \bibinfo{person}{Jinlan Fu}, \bibinfo{person}{Zhengbao Jiang}, \bibinfo{person}{Hiroaki Hayashi}, {and} \bibinfo{person}{Graham Neubig}.} \bibinfo{year}{2023}\natexlab{}.
\newblock \showarticletitle{Pre-train, prompt, and predict: A systematic survey of prompting methods in natural language processing}.
\newblock \bibinfo{journal}{\emph{Comput. Surveys}} \bibinfo{volume}{55}, \bibinfo{number}{9} (\bibinfo{year}{2023}), \bibinfo{pages}{1--35}.
\newblock


\bibitem[Ni et~al\mbox{.}(2024)]%
        {ni2024casmodatest}
\bibfield{author}{\bibinfo{person}{Chao Ni}, \bibinfo{person}{Xiaoya Wang}, \bibinfo{person}{Liushan Chen}, \bibinfo{person}{Dehai Zhao}, \bibinfo{person}{Zhengong Cai}, \bibinfo{person}{Shaohua Wang}, {and} \bibinfo{person}{Xiaohu Yang}.} \bibinfo{year}{2024}\natexlab{}.
\newblock \showarticletitle{CasModaTest: A Cascaded and Model-agnostic Self-directed Framework for Unit Test Generation}.
\newblock \bibinfo{journal}{\emph{arXiv preprint arXiv:2406.15743}} (\bibinfo{year}{2024}).
\newblock


\bibitem[Ni et~al\mbox{.}(2023)]%
        {ni2023distinguishing}
\bibfield{author}{\bibinfo{person}{Chao Ni}, \bibinfo{person}{Xin Yin}, \bibinfo{person}{Kaiwen Yang}, \bibinfo{person}{Dehai Zhao}, \bibinfo{person}{Zhenchang Xing}, {and} \bibinfo{person}{Xin Xia}.} \bibinfo{year}{2023}\natexlab{}.
\newblock \showarticletitle{Distinguishing Look-Alike Innocent and Vulnerable Code by Subtle Semantic Representation Learning and Explanation}. In \bibinfo{booktitle}{\emph{Proceedings of the 31st ACM Joint European Software Engineering Conference and Symposium on the Foundations of Software Engineering}}. \bibinfo{pages}{1611--1622}.
\newblock


\bibitem[Oliveira et~al\mbox{.}(2018)]%
        {oliveira2018mapping}
\bibfield{author}{\bibinfo{person}{Carlos Oliveira}, \bibinfo{person}{Aldeida Aleti}, \bibinfo{person}{Lars Grunske}, {and} \bibinfo{person}{Kate Smith-Miles}.} \bibinfo{year}{2018}\natexlab{}.
\newblock \showarticletitle{Mapping the effectiveness of automated test suite generation techniques}.
\newblock \bibinfo{journal}{\emph{IEEE Transactions on Reliability}} \bibinfo{volume}{67}, \bibinfo{number}{3} (\bibinfo{year}{2018}), \bibinfo{pages}{771--785}.
\newblock


\bibitem[OpenAI(2022)]%
        {openai2022chatgpt}
\bibfield{author}{\bibinfo{person}{OpenAI}.} \bibinfo{year}{2022}\natexlab{}.
\newblock \bibinfo{title}{ChatGPT: Optimizing Language Models for Dialogue. (2022)}.
\newblock \bibinfo{howpublished}{\url{https://openai.com/blog/chatgpt/}}.
\newblock


\bibitem[Ouyang et~al\mbox{.}(2022)]%
        {ouyang2022training}
\bibfield{author}{\bibinfo{person}{Long Ouyang}, \bibinfo{person}{Jeffrey Wu}, \bibinfo{person}{Xu Jiang}, \bibinfo{person}{Diogo Almeida}, \bibinfo{person}{Carroll Wainwright}, \bibinfo{person}{Pamela Mishkin}, \bibinfo{person}{Chong Zhang}, \bibinfo{person}{Sandhini Agarwal}, \bibinfo{person}{Katarina Slama}, \bibinfo{person}{Alex Ray}, {et~al\mbox{.}}} \bibinfo{year}{2022}\natexlab{}.
\newblock \showarticletitle{Training language models to follow instructions with human feedback}.
\newblock \bibinfo{journal}{\emph{Advances in Neural Information Processing Systems}}  \bibinfo{volume}{35} (\bibinfo{year}{2022}), \bibinfo{pages}{27730--27744}.
\newblock


\bibitem[Pacheco and Ernst(2007)]%
        {pacheco2007randoop}
\bibfield{author}{\bibinfo{person}{Carlos Pacheco} {and} \bibinfo{person}{Michael~D Ernst}.} \bibinfo{year}{2007}\natexlab{}.
\newblock \showarticletitle{Randoop: feedback-directed random testing for Java}. In \bibinfo{booktitle}{\emph{Companion to the 22nd ACM SIGPLAN conference on Object-oriented programming systems and applications companion}}. \bibinfo{pages}{815--816}.
\newblock


\bibitem[Panichella et~al\mbox{.}(2015)]%
        {panichella2015reformulating}
\bibfield{author}{\bibinfo{person}{Annibale Panichella}, \bibinfo{person}{Fitsum~Meshesha Kifetew}, {and} \bibinfo{person}{Paolo Tonella}.} \bibinfo{year}{2015}\natexlab{}.
\newblock \showarticletitle{Reformulating branch coverage as a many-objective optimization problem}. In \bibinfo{booktitle}{\emph{2015 IEEE 8th international conference on software testing, verification and validation (ICST)}}. IEEE, \bibinfo{pages}{1--10}.
\newblock


\bibitem[Panichella et~al\mbox{.}(2017)]%
        {panichella2017automated}
\bibfield{author}{\bibinfo{person}{Annibale Panichella}, \bibinfo{person}{Fitsum~Meshesha Kifetew}, {and} \bibinfo{person}{Paolo Tonella}.} \bibinfo{year}{2017}\natexlab{}.
\newblock \showarticletitle{Automated test case generation as a many-objective optimisation problem with dynamic selection of the targets}.
\newblock \bibinfo{journal}{\emph{IEEE Transactions on Software Engineering}} \bibinfo{volume}{44}, \bibinfo{number}{2} (\bibinfo{year}{2017}), \bibinfo{pages}{122--158}.
\newblock


\bibitem[Paszke et~al\mbox{.}(2019)]%
        {pytorch}
\bibfield{author}{\bibinfo{person}{Adam Paszke}, \bibinfo{person}{Sam Gross}, \bibinfo{person}{Francisco Massa}, \bibinfo{person}{Adam Lerer}, \bibinfo{person}{James Bradbury}, \bibinfo{person}{Gregory Chanan}, \bibinfo{person}{Trevor Killeen}, \bibinfo{person}{Zeming Lin}, \bibinfo{person}{Natalia Gimelshein}, \bibinfo{person}{Luca Antiga}, \bibinfo{person}{Alban Desmaison}, \bibinfo{person}{Andreas Kopf}, \bibinfo{person}{Edward Yang}, \bibinfo{person}{Zachary DeVito}, \bibinfo{person}{Martin Raison}, \bibinfo{person}{Alykhan Tejani}, \bibinfo{person}{Sasank Chilamkurthy}, \bibinfo{person}{Benoit Steiner}, \bibinfo{person}{Lu Fang}, \bibinfo{person}{Junjie Bai}, {and} \bibinfo{person}{Soumith Chintala}.} \bibinfo{year}{2019}\natexlab{}.
\newblock \showarticletitle{PyTorch: An Imperative Style, High-Performance Deep Learning Library}.
\newblock In \bibinfo{booktitle}{\emph{Advances in Neural Information Processing Systems 32}}. \bibinfo{publisher}{Curran Associates, Inc.}, \bibinfo{pages}{8024--8035}.
\newblock
\urldef\tempurl%
\url{http://papers.neurips.cc/paper/9015-pytorch-an-imperative-style-high-performance-deep-learning-library.pdf}
\showURL{%
\tempurl}


\bibitem[Radford et~al\mbox{.}(2018)]%
        {radford2018improving}
\bibfield{author}{\bibinfo{person}{Alec Radford}, \bibinfo{person}{Karthik Narasimhan}, \bibinfo{person}{Tim Salimans}, \bibinfo{person}{Ilya Sutskever}, {et~al\mbox{.}}} \bibinfo{year}{2018}\natexlab{}.
\newblock \showarticletitle{Improving language understanding by generative pre-training}.
\newblock  (\bibinfo{year}{2018}).
\newblock


\bibitem[Rao et~al\mbox{.}(2023)]%
        {rao2023cat}
\bibfield{author}{\bibinfo{person}{Nikitha Rao}, \bibinfo{person}{Kush Jain}, \bibinfo{person}{Uri Alon}, \bibinfo{person}{Claire Le~Goues}, {and} \bibinfo{person}{Vincent~J Hellendoorn}.} \bibinfo{year}{2023}\natexlab{}.
\newblock \showarticletitle{CAT-LM training language models on aligned code and tests}. In \bibinfo{booktitle}{\emph{2023 38th IEEE/ACM International Conference on Automated Software Engineering (ASE)}}. IEEE, \bibinfo{pages}{409--420}.
\newblock


\bibitem[Rozi{\`e}re et~al\mbox{.}(2023)]%
        {roziere2023code}
\bibfield{author}{\bibinfo{person}{Baptiste Rozi{\`e}re}, \bibinfo{person}{Jonas Gehring}, \bibinfo{person}{Fabian Gloeckle}, \bibinfo{person}{Sten Sootla}, \bibinfo{person}{Itai Gat}, \bibinfo{person}{Xiaoqing~Ellen Tan}, \bibinfo{person}{Yossi Adi}, \bibinfo{person}{Jingyu Liu}, \bibinfo{person}{Tal Remez}, \bibinfo{person}{J{\'e}r{\'e}my Rapin}, {et~al\mbox{.}}} \bibinfo{year}{2023}\natexlab{}.
\newblock \showarticletitle{Code llama: Open foundation models for code}.
\newblock \bibinfo{journal}{\emph{arXiv preprint arXiv:2308.12950}} (\bibinfo{year}{2023}).
\newblock


\bibitem[Ryan et~al\mbox{.}(2024)]%
        {ryan2024code}
\bibfield{author}{\bibinfo{person}{Gabriel Ryan}, \bibinfo{person}{Siddhartha Jain}, \bibinfo{person}{Mingyue Shang}, \bibinfo{person}{Shiqi Wang}, \bibinfo{person}{Xiaofei Ma}, \bibinfo{person}{Murali~Krishna Ramanathan}, {and} \bibinfo{person}{Baishakhi Ray}.} \bibinfo{year}{2024}\natexlab{}.
\newblock \showarticletitle{Code-aware prompting: A study of coverage-guided test generation in regression setting using llm}.
\newblock \bibinfo{journal}{\emph{Proceedings of the ACM on Software Engineering}} \bibinfo{volume}{1}, \bibinfo{number}{FSE} (\bibinfo{year}{2024}), \bibinfo{pages}{951--971}.
\newblock


\bibitem[Saes(2018)]%
        {saes2018unit}
\bibfield{author}{\bibinfo{person}{Laurence Saes}.} \bibinfo{year}{2018}\natexlab{}.
\newblock \showarticletitle{Unit test generation using machine learning}.
\newblock \bibinfo{journal}{\emph{Universiteit van Amsterdamg}} (\bibinfo{year}{2018}).
\newblock


\bibitem[Sch{\"a}fer et~al\mbox{.}(2023)]%
        {schafer2023empirical}
\bibfield{author}{\bibinfo{person}{Max Sch{\"a}fer}, \bibinfo{person}{Sarah Nadi}, \bibinfo{person}{Aryaz Eghbali}, {and} \bibinfo{person}{Frank Tip}.} \bibinfo{year}{2023}\natexlab{}.
\newblock \showarticletitle{An empirical evaluation of using large language models for automated unit test generation}.
\newblock \bibinfo{journal}{\emph{IEEE Transactions on Software Engineering}} (\bibinfo{year}{2023}).
\newblock


\bibitem[Shieh(2023)]%
        {shieh2023best}
\bibfield{author}{\bibinfo{person}{Jessica Shieh}.} \bibinfo{year}{2023}\natexlab{}.
\newblock \showarticletitle{Best practices for prompt engineering with OpenAI API}.
\newblock \bibinfo{journal}{\emph{OpenAI, February https://help.openai. com/en/articles/6654000-best-practices-for-prompt-engineering-with-openai-api}} (\bibinfo{year}{2023}).
\newblock


\bibitem[Shin et~al\mbox{.}(2024)]%
        {shin2024domain}
\bibfield{author}{\bibinfo{person}{Jiho Shin}, \bibinfo{person}{Sepehr Hashtroudi}, \bibinfo{person}{Hadi Hemmati}, {and} \bibinfo{person}{Song Wang}.} \bibinfo{year}{2024}\natexlab{}.
\newblock \showarticletitle{Domain Adaptation for Code Model-Based Unit Test Case Generation}. In \bibinfo{booktitle}{\emph{Proceedings of the 33rd ACM SIGSOFT International Symposium on Software Testing and Analysis}}. \bibinfo{pages}{1211--1222}.
\newblock


\bibitem[Tufano et~al\mbox{.}(2022)]%
        {tufano2022methods2test}
\bibfield{author}{\bibinfo{person}{Michele Tufano}, \bibinfo{person}{Shao~Kun Deng}, \bibinfo{person}{Neel Sundaresan}, {and} \bibinfo{person}{Alexey Svyatkovskiy}.} \bibinfo{year}{2022}\natexlab{}.
\newblock \showarticletitle{Methods2Test: A dataset of focal methods mapped to test cases}. In \bibinfo{booktitle}{\emph{Proceedings of the 19th International Conference on Mining Software Repositories}}. \bibinfo{pages}{299--303}.
\newblock


\bibitem[Tufano et~al\mbox{.}(2020)]%
        {tufano2020unit}
\bibfield{author}{\bibinfo{person}{Michele Tufano}, \bibinfo{person}{Dawn Drain}, \bibinfo{person}{Alexey Svyatkovskiy}, \bibinfo{person}{Shao~Kun Deng}, {and} \bibinfo{person}{Neel Sundaresan}.} \bibinfo{year}{2020}\natexlab{}.
\newblock \showarticletitle{Unit test case generation with transformers and focal context}.
\newblock \bibinfo{journal}{\emph{arXiv preprint arXiv:2009.05617}} (\bibinfo{year}{2020}).
\newblock


\bibitem[Vaswani et~al\mbox{.}(2017)]%
        {vaswani2017attention}
\bibfield{author}{\bibinfo{person}{Ashish Vaswani}, \bibinfo{person}{Noam Shazeer}, \bibinfo{person}{Niki Parmar}, \bibinfo{person}{Jakob Uszkoreit}, \bibinfo{person}{Llion Jones}, \bibinfo{person}{Aidan~N Gomez}, \bibinfo{person}{{\L}ukasz Kaiser}, {and} \bibinfo{person}{Illia Polosukhin}.} \bibinfo{year}{2017}\natexlab{}.
\newblock \showarticletitle{Attention is all you need}.
\newblock \bibinfo{journal}{\emph{Advances in neural information processing systems}}  \bibinfo{volume}{30} (\bibinfo{year}{2017}).
\newblock


\bibitem[Wang et~al\mbox{.}(2023)]%
        {wang2023nxtunit}
\bibfield{author}{\bibinfo{person}{Siwei Wang}, \bibinfo{person}{Xue Mao}, \bibinfo{person}{Ziguang Cao}, \bibinfo{person}{Yujun Gao}, \bibinfo{person}{Qucheng Shen}, {and} \bibinfo{person}{Chao Peng}.} \bibinfo{year}{2023}\natexlab{}.
\newblock \showarticletitle{NxtUnit: Automated Unit Test Generation for Go}. In \bibinfo{booktitle}{\emph{Proceedings of the 27th International Conference on Evaluation and Assessment in Software Engineering}}. \bibinfo{pages}{176--179}.
\newblock


\bibitem[Wang et~al\mbox{.}(2021)]%
        {wang2021codet5}
\bibfield{author}{\bibinfo{person}{Yue Wang}, \bibinfo{person}{Weishi Wang}, \bibinfo{person}{Shafiq Joty}, {and} \bibinfo{person}{Steven~CH Hoi}.} \bibinfo{year}{2021}\natexlab{}.
\newblock \showarticletitle{Codet5: Identifier-aware unified pre-trained encoder-decoder models for code understanding and generation}.
\newblock \bibinfo{journal}{\emph{arXiv preprint arXiv:2109.00859}} (\bibinfo{year}{2021}).
\newblock


\bibitem[Wei et~al\mbox{.}(2023)]%
        {wei2023magicoder}
\bibfield{author}{\bibinfo{person}{Yuxiang Wei}, \bibinfo{person}{Zhe Wang}, \bibinfo{person}{Jiawei Liu}, \bibinfo{person}{Yifeng Ding}, {and} \bibinfo{person}{Lingming Zhang}.} \bibinfo{year}{2023}\natexlab{}.
\newblock \showarticletitle{Magicoder: Source code is all you need}.
\newblock \bibinfo{journal}{\emph{arXiv preprint arXiv:2312.02120}} (\bibinfo{year}{2023}).
\newblock


\bibitem[Xia et~al\mbox{.}(2023)]%
        {xia2023automated}
\bibfield{author}{\bibinfo{person}{Chunqiu~Steven Xia}, \bibinfo{person}{Yuxiang Wei}, {and} \bibinfo{person}{Lingming Zhang}.} \bibinfo{year}{2023}\natexlab{}.
\newblock \showarticletitle{Automated program repair in the era of large pre-trained language models}. In \bibinfo{booktitle}{\emph{Proceedings of the 45th International Conference on Software Engineering (ICSE 2023). Association for Computing Machinery}}.
\newblock


\bibitem[Xia and Zhang(2023)]%
        {xia2023keep}
\bibfield{author}{\bibinfo{person}{Chunqiu~Steven Xia} {and} \bibinfo{person}{Lingming Zhang}.} \bibinfo{year}{2023}\natexlab{}.
\newblock \showarticletitle{Keep the Conversation Going: Fixing 162 out of 337 bugs for \$0.42 each using ChatGPT}.
\newblock \bibinfo{journal}{\emph{arXiv preprint arXiv:2304.00385}} (\bibinfo{year}{2023}).
\newblock


\bibitem[Yin et~al\mbox{.}(2024b)]%
        {yin2024rectifier}
\bibfield{author}{\bibinfo{person}{Xin Yin}, \bibinfo{person}{Chao Ni}, \bibinfo{person}{Tien~N Nguyen}, \bibinfo{person}{Shaohua Wang}, {and} \bibinfo{person}{Xiaohu Yang}.} \bibinfo{year}{2024}\natexlab{b}.
\newblock \showarticletitle{Rectifier: Code Translation with Corrector via LLMs}.
\newblock \bibinfo{journal}{\emph{arXiv preprint arXiv:2407.07472}} (\bibinfo{year}{2024}).
\newblock


\bibitem[Yin et~al\mbox{.}(2024a)]%
        {yin2024multitask}
\bibfield{author}{\bibinfo{person}{Xin Yin}, \bibinfo{person}{Chao Ni}, {and} \bibinfo{person}{Shaohua Wang}.} \bibinfo{year}{2024}\natexlab{a}.
\newblock \showarticletitle{Multitask-based evaluation of open-source llm on software vulnerability}.
\newblock \bibinfo{journal}{\emph{IEEE Transactions on Software Engineering}} (\bibinfo{year}{2024}).
\newblock


\bibitem[Yin et~al\mbox{.}(2024c)]%
        {yin2024thinkrepair}
\bibfield{author}{\bibinfo{person}{Xin Yin}, \bibinfo{person}{Chao Ni}, \bibinfo{person}{Shaohua Wang}, \bibinfo{person}{Zhenhao Li}, \bibinfo{person}{Limin Zeng}, {and} \bibinfo{person}{Xiaohu Yang}.} \bibinfo{year}{2024}\natexlab{c}.
\newblock \showarticletitle{Thinkrepair: Self-directed automated program repair}. In \bibinfo{booktitle}{\emph{Proceedings of the 33rd ACM SIGSOFT International Symposium on Software Testing and Analysis}}. \bibinfo{pages}{1274--1286}.
\newblock


\bibitem[Yin et~al\mbox{.}(2024d)]%
        {yin2024you}
\bibfield{author}{\bibinfo{person}{Xin Yin}, \bibinfo{person}{Chao Ni}, \bibinfo{person}{Xiaodan Xu}, {and} \bibinfo{person}{Xiaohu Yang}.} \bibinfo{year}{2024}\natexlab{d}.
\newblock \showarticletitle{What You See Is What You Get: Attention-based Self-guided Automatic Unit Test Generation}.
\newblock \bibinfo{journal}{\emph{arXiv preprint arXiv:2412.00828}} (\bibinfo{year}{2024}).
\newblock


\bibitem[Yuan et~al\mbox{.}(2024)]%
        {yuan2024evaluating}
\bibfield{author}{\bibinfo{person}{Zhiqiang Yuan}, \bibinfo{person}{Mingwei Liu}, \bibinfo{person}{Shiji Ding}, \bibinfo{person}{Kaixin Wang}, \bibinfo{person}{Yixuan Chen}, \bibinfo{person}{Xin Peng}, {and} \bibinfo{person}{Yiling Lou}.} \bibinfo{year}{2024}\natexlab{}.
\newblock \showarticletitle{Evaluating and improving chatgpt for unit test generation}.
\newblock \bibinfo{journal}{\emph{Proceedings of the ACM on Software Engineering}} \bibinfo{volume}{1}, \bibinfo{number}{FSE} (\bibinfo{year}{2024}), \bibinfo{pages}{1703--1726}.
\newblock


\bibitem[Yuan et~al\mbox{.}(2023)]%
        {yuan2023no}
\bibfield{author}{\bibinfo{person}{Zhiqiang Yuan}, \bibinfo{person}{Yiling Lou}, \bibinfo{person}{Mingwei Liu}, \bibinfo{person}{Shiji Ding}, \bibinfo{person}{Kaixin Wang}, \bibinfo{person}{Yixuan Chen}, {and} \bibinfo{person}{Xin Peng}.} \bibinfo{year}{2023}\natexlab{}.
\newblock \showarticletitle{No more manual tests? evaluating and improving chatgpt for unit test generation}.
\newblock \bibinfo{journal}{\emph{arXiv preprint arXiv:2305.04207}} (\bibinfo{year}{2023}).
\newblock


\bibitem[Ziegler et~al\mbox{.}(2019)]%
        {ziegler2019fine}
\bibfield{author}{\bibinfo{person}{Daniel~M Ziegler}, \bibinfo{person}{Nisan Stiennon}, \bibinfo{person}{Jeffrey Wu}, \bibinfo{person}{Tom~B Brown}, \bibinfo{person}{Alec Radford}, \bibinfo{person}{Dario Amodei}, \bibinfo{person}{Paul Christiano}, {and} \bibinfo{person}{Geoffrey Irving}.} \bibinfo{year}{2019}\natexlab{}.
\newblock \showarticletitle{Fine-tuning language models from human preferences}.
\newblock \bibinfo{journal}{\emph{arXiv preprint arXiv:1909.08593}} (\bibinfo{year}{2019}).
\newblock


\end{thebibliography}

\end{document}